\documentclass[manuscript]{aastex}
\newcommand{\noprint}[1]{}

\shorttitle{Herschel Observations of Active Galaxies}
\shortauthors{Mel\'endez et al.}

\begin{document}


\title{Herschel Far-Infrared Photometry of the Swift Burst Alert Telescope Active Galactic Nuclei Sample of the Local Universe. I. PACS Observations\thanks{{\it Herschel} is an ESA space observatory with science instruments provided by European-led Principal Investigator consortia and with important participation from NASA.}}


\author{M. Mel\'endez; R. F. Mushotzky \and T. T. Shimizu}
\affil{Astronomy Department, University of Maryland, Stadium Dr. , College Park, MD, 20742, USA}
\email{marcio@astro.umd.edu}
\author{A. J. Barger\altaffilmark{1,2}}
\affil{Department of Astronomy, University of Wisconsin-Madison, 475 N. Charter Street, Madison, WI 53706, USA}
\and 
\author{L. L. Cowie}
\affil{Institute for Astronomy, University of Hawaii, 2680 Woodlawn Drive, Honolulu, HI 96822, USA}

\altaffiltext{1}{Department of Physics and Astronomy, University of Hawaii, 2505 Correa Road, Honolulu, HI 96822, USA}
\altaffiltext{2}{Institute for Astronomy, University of Hawaii, 2680 Woodlawn Drive, Honolulu, HI 96822, USA}





\begin{abstract}
Far-Infrared (FIR) photometry from the the Photodetector Array Camera and Spectrometer (PACS)   on the {\it Herschel} Space Observatory is presented for 313 nearby, hard X-ray selected galaxies from the 58-month  {\it Swift}  Burst Alert Telescope (BAT) Active Galactic catalog. The present data do not distinguish between the FIR  luminosity distributions at 70 and 160~$\micron$  for  Seyfert 1 and Seyfert 2 galaxies. This result suggests that if the FIR  emission is from  the nuclear obscuring material surrounding the accretion disk,  then it emits isotropically, independent of orientation. Alternatively, a significant fraction of the 70 and 160~$\micron$ could be from star formation, independent of AGN type. Using a non-parametric test for partial correlation with censored data, we find  a statistically significant correlation between the AGN intrinsic power (in the 14-195 keV band ) and the FIR emission at 70 and 160 $\micron$ for Seyfert 1 galaxies. We find no correlation between the 14-195 keV and FIR luminosities in Seyfert 2 galaxies.  The observed correlations  suggest two possible scenarios: (i) if we assume  that the FIR luminosity is a good tracer of star formation,  then there is a connection between star formation and the AGN at sub-kiloparsec scales, or (ii) dust heated by the  AGN has a statistically significant  contribution to the FIR emission.  Using a Spearman rank-order analysis, the 14-195 keV luminosities for the Seyfert 1 and 2 galaxies are weakly statistically correlated with the F$_{70}$/F$_{160}$  ratios. 
\end{abstract}


\section{Introduction}

In the past few years there  has been an incredible amount of information regarding the infrared view of our dusty Universe. With the advent of very successful space missions such as {\it Spitzer} \citep{2004ApJS..154....1W}, NASA's {\it  Wide-field Infrared Survey Explorer} \citep[WISE, ][]{2010AJ....140.1868W}, {\it AKARI} \citep{2007PASJ...59S.369M}, {\it Plank} \citep{2011A&A...536A...1P} and {\it Herschel} \citep{2010A&A...518L...1P}, the mid-infrared (MIR)  and far-infrared (FIR) is beginning to provide a unique perspective into the evolution of galaxies via dust-based star formation rate (SFR) indicators. However, these SFR indicators are not free from systematic uncertainties primarily due to the nature of dust in the galaxy by trapping the  starlight and re-emitting (a fraction) in the infrared. Because of this, each MIR and FIR band can be associated with a different dust component,  different spatial distributions and  different stellar age populations \citep[see][for a review]{2012ARA&A..50..531K}. For  these reasons, monochromatic infrared emission may be restricted, as a reliable SFR indicator, to large samples of galaxies for  statistically  significant studies \citep[e.g.,][]{2010ApJ...714.1256C}. On the other hand,   SFR indicators that rely on the total FIR emission are subject to  uncertainties  because they  are very susceptible to the shape of the MIR to FIR spectral energy distribution \citep[e.g.,][]{1998ARA&A..36..189K}. Despite these considerations and given the plethora of new and more sensitive infrared surveys, these SFR metrics  are widely used. However, little consideration is given to the possible  contribution from dust heated by a non-stellar ionization source, i.e., the active galactic nucleus (AGN) in the center of the galaxy.

Following the unified model of AGN \citep[where the various  types are explained solely by a viewing angle difference; see ][for details]{1993ARA&A..31..473A},  the hot accretion disk around the galaxies  supermassive black hole is surrounded by a dusty, molecular torus. In recent years, there has been considerable  effort to characterize the nature of this obscuring material both theoretically  and observationally. Several models have been proposed to explain  the observed and inferred properties of the torus, including a   smooth, continuous, geometrically and optically thick  dusty torus \citep[][]{1988ApJ...329..702K,2004A&A...426..445B,2006MNRAS.366..767F}, full radiative transfer clumpy torus  models \citep[e.g.,][]{2008ApJ...685..147N,2008A&A...482...67S,2009ApJ...705..298M,2010A&A...523A..27H,2012MNRAS.420.2756S} and   clumpy  winds structures  originating from the accretion disk \citep[e.g.,][]{2006ApJ...648L.101E}. However, within their uncertainties (e.g., size of the torus, inclination, dust distribution function, etc), almost all these models predict some (non-negligible) torus contribution at the MIR  and FIR  wavelengths where dust-based SFRs are commonly used under the assumption that the FIR luminosity is attributed solely to star formation \citep[e.g.,][]{2006MNRAS.366..767F,2011MNRAS.414.1082M}. Thus, it is of the utmost importance   to determine the AGN contribution to the FIR emission.

To investigate  the AGN contribution to the FIR emission, it is important to start with a complete and  unbiased sample of local AGN at $z <$ 0.05, where we can use {\it Herschel's} unique angular resolution to spatially resolve the  FIR emission.  Given its high energy band selection, 14 to  195~keV, the {\it Swift}/Burst Alert Telescope (BAT)  survey of AGN represents such a sample,  because  it is unbiased to Compton thin AGN, e.g., sources with Hydrogen column densities of less than a few times $10^{24}$~${\rm cm^{-2}}$. In addition, the BAT AGN survey is  not sensitive to stellar activity in the host galaxy,  because  star formation has a negligible contribution at these hard X-ray energies. Thus,  the BAT  sample can provide  a  unique perspective into the AGN and star formation contributions at the FIR wavelengths, where the dusty torus and the current star formation are the competing effects to  heat the dust. For this purpose,  we performed a statistical study of the correlations between the FIR luminosities  observed by the {\it Herschel} observatory and the hard X-ray luminosities for more than  300 local BAT AGN. In Section~2-3 we present details on the sample selection, {\it Herschel} Observations and data processing. Section~4 shows the FIR properties of the BAT sample. Section~5 discusses the observed correlations between the hard X-ray and  FIR emission.  In Section~6 we present the FIR colors of the BAT sample and their  implications for some of the FIR predictions of torus models. Section~7 shows the comparison for the FIR colors between the BAT AGN and normal, star forming galaxies. Finally, Section~8 lists the main conclusions of this work.  In order to calculate luminosities  we assumed a flat universe with a Hubble constant H$_o$ = 71 km s$^{-1}$ Mpc$^{-1}$, $\Omega_\Lambda$ = 0.73, and $\Omega_M$ = 0.27, with redshift values taken from NASA's ExtraGalactic Database (NED), except for sources with redshift values  $z <$ 0.01, where distances are taken from The Extragalactic Distance Database \citep[EDD; ][]{1988ngc..book.....T,2009AJ....138..323T}.

\section{Sample}
The  sample presented in this work  was selected from the the low-redshift ($z<0.05$) 58-month Swift-BAT survey with a  median redshift of $ z \sim$0.025\footnote{http://swift.gsfc.nasa.gov/results/bs58mon/}. The   58-month Swift-BAT is an almost  uniform hard X-ray all-sky survey and reaches a flux level of 1.1$\times$10$^{-11}$ ergs sec$^{-1}$ cm$^{-2}$ over 50\% of the sky and 1.48$\times$10$^{-11}$ ergs sec$^{-1}$ cm$^{-2}$ over 90\% of the sky \citep{2013ApJS..207...19B}. Source identifications are based primarily on the X-ray imaging data and a correlation with optical images and catalogs. In some cases, the identifications are based on positional coincidences with previously known AGN. The main advantage of the BAT AGN sample is that  the selection process is completely independent of optical, IR or radio properties of the host galaxy. Our final sample of galaxies includes 149 Seyfert 1 galaxies (1/1.2/1.5), 157 Seyfert 2 galaxies (1.8/1.9/2.0), 6 LINERs and 1 unclassified Seyfert galaxy, ESO 464-G016 \citep{2010A&A...518A..10V}. One must note that some of the Seyfert galaxies have dual classifications, see Table~\ref{table1}. For the purpose of grouping galaxies through the rest of the paper, we take the Seyfert classification as the primary type.

\section{Observations and  Data Processing}


The 58-month BAT sample was observed by  the Photodetector Array Camera and Spectrometer \citep[PACS, ][]{2010A&A...518L...2P}  on the {\it Herschel} Space Observatory. The vast majority of the BAT AGN presented in this work are from our  cycle 1 open-time program (OT1\_rmushot\_1, PI: R. Mushotzky) with a total of 291 sources.  For the sake of completeness we included 22 BAT sources from different programs publicly  available from the Herschel science archive (HSA), see Table~\ref{table1} for details. From this the total number of BAT AGN sources in our sample is 313. For the sources obtained through our OT1 program the PACS imaging for the blue 70~$\micron$  (60-85  $\micron$) and red 160 $\micron$  (130-210 $\micron$) band  was obtained simultaneously in scan mode along two  scan map position angles at  70 and 110 degrees. Each orientation angle was scanned with a   medium scan speed of 20\arcsec~ s$^{-1}$,   2 scan legs of 3.0\arcmin~ length with 5.0\arcsec~ scan leg separation and a repetition factor of 1. The total time per observation was 52 s. From our OT1 program the galaxies  II SZ 010, Mrk 290, PG 2304+042 and Mrk 841 have a different configuration with 10 scan legs of 3.0\arcmin~ length with 4.0\arcsec~ scan leg separation and a repetition factor of 1 with a total time per observation of 276 s.

For the PACS data reduction we use the Herschel Interactive Processing Environment \citep[HIPE,][]{2010ASPC..434..139O} version 8.0. The ``Level 0" observations (raw data)  were processed through the standard pipeline procedure to convert from Level 0 to Level 1 data. This  procedure includes the extraction of the calibration tree needed for the data processing, correction for electronic crosstalk, application of the flat-field correction and finally  deglitching and conversion from  Volts to Jy/pixel. In order to correct for bolometer drift (low frequency noise), both thermal and non-thermal (uncorrelated noise), and to create the final maps from the Level 1 data,   we used the algorithm implemented in {\it Scanamorphos} \citep[v19.0,][]{2013PASP..125.1126R},  which makes use of   the redundancy built in the observations to derive the brightness drifts. Because of this,  Scanamorphos is independent of any pre-defined noise model because it  relies  on the fact that each portion of the sky is scanned  by multiple bolometers at different times. All final maps have a pixel size of $\sim$1/4 of the point spread function (PSF) full width at half-maximum (FWHM), i.e., 1.4\arcsec~ at 70 $\micron$  and 2.85\arcsec~ at 160 $\micron$.  {\it Scanamorphos} also produces an error and weight map. The error map is defined as the error on the mean brightness in each pixel. It is built using the weighted variance because weights are used for the projection of the final map. The error map does not include any error propagation associated with the different steps performed on the pipeline.  On the other hand,  the weighted map is built by co-adding the weights and is normalized by the average of the weights. See Figure~\ref{mosaic} of the Appendix for a sample of the final maps. All the images are available in electronic form in the on-line version. At the median redshift for the sample, 1\arcsec~ represents  $\sim$500~pc. Hence, for example, PACS~70$\micron$ PSF will sample about 2.8~kpc at $ z \sim$0.025.

PACS fluxes are measured using a combination of circular and elliptical apertures for sources that are visually identified  as point-like  (and/or relatively point-like) and extended sources, respectively. The apertures  are chosen by eye to contain all of the observed emission at each wavelength. In addition, background subtraction is performed locally with a circular or elliptical annulus around the source. For point-like sources the background annulus was set to be 20\arcsec~ to 25\arcsec~ and 24\arcsec~ to 28\arcsec~ in radius for the blue and red camera, respectively. For extended sources,  the background annulus was set to encompass a clean, uncontaminated sky region close to the source. Table~\ref{table1} shows the different aperture sizes used in this work. 

Aperture corrections are applied to the background subtracted fluxes. For point-like sources we applied the correction outlined in Table 15 of the technical report PICC-ME-TN-037\footnote{https://nhscdmz2.ipac.caltech.edu/sc/uploads/Pacs/PSF\_aperture\_corrections\_FM6.txt}. To derive aperture corrections for extended sources we used  higher resolution images  from {\it Spitzer}/IRAC 8.0 $\micron$ and  measured the  total flux with the same aperture employed for the PACS analysis. Then we convolved the same image  with the right kernel to bring it to the PACS resolution\footnote{http://dirty.as.arizona.edu/~kgordon/mips/conv\_psfs/conv\_psfs.html} \citep{2008ApJ...682..336G} and remeasured  the flux in the same aperture. The ratio of the unconvolved to the  convolved (the same PSF as the {\it Herschel} PACS)  flux is used as an estimate of the aperture correction. From the list  of extended sources, there are 53 BAT AGN  sources with {\it Spitzer}/IRAC 8.0 $\micron$ images available  through The Spitzer Enhanced Imaging Products archive. For these sources we find the aperture corrections to be very small, with values not greater than 1.03 ($< 3\%$) and with a mean value very close to unity. Therefore, for  large, extended sources without higher resolution images, we apply no correction,  so we added a 3\% flux uncertainty. Caution must be taken, because aperture corrections derived from this empirical method assume that the spatial distribution  is the same at PACS (FIR) and IRAC (MIR) wavelengths.

The total uncertainty for the integrated photometry is a combination of the error on the mean brightness in each pixel added in quadrature within the source aperture (the error map produced by Scanamorphos), the standard deviation of all the pixels  in the background aperture and the PACS photometer flux calibration accuracy. In some cases, the dominant source of error is due to   background fluctuations especially in sources contaminated by cirrus (see Table~\ref{table1}); however, almost all of our sources have a clean, flat  extragalactic field. The PACS calibration uncertainties are $\sigma_{cal}$, ∼5\%, according to Version 2.4  of the PACS Observer's Manual\footnote{http://herschel.esac.esa.int/Docs/PACS/html/pacs\_om.html}. Finally,  we take as an upper-limit five times the total uncertainty for the integrated photometry (5$\sigma$). Table~\ref{table1} presents the spatially integrated flux densities for all 313  galaxies for PACS blue and red  photometric bands. The tabulated flux densities include aperture corrections.   No reddening and color corrections have been applied to the data in Table~\ref{table1}. From this, 295 and 260 sources are detected  by PACS at 70 and 160~$\micron$, implying a detection rate of 94\% and 83\%, respectively. Only two sources, namely, MCG-01-09-045 and UGC03995A,  are detected at PACS 160~$\micron$ with no detections at 70~$\micron$.

\begin{figure}
\epsscale{.50}
\plotone{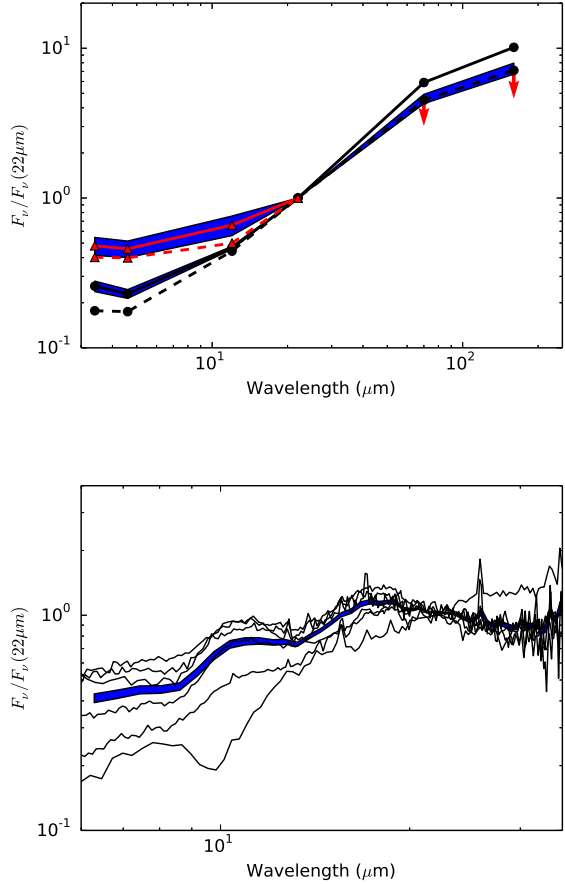}
\caption{Upper panel: comparison of the observed spectral energy distribution, normalized to the flux in the {\it WISE} 22 $\micron$ band, for the entire BAT sample (circles) and that of the PACS undetected sources (triangles, BAT sources undetected in both PACS band, see Table~\ref{table1}). The mean and median of the entire sample are represented by the  solid and dashed black lines, respectively. Similarly,  the mean and median of the PACS undetected sources are shown as solid and dashed red lines, respectively. The  blue shading indicates the standard error of the mean. Lower panel: {\it Spitzer} low-resolution  spectra for a sample of PACS undetected sources observed by {\it Spitzer}, namely, ESO~121-IG028, Mrk~352, Mrk~50, PG2304+042, SBS1301+540 and UM~614.  The individual spectra have been normalized to the flux at 22~$\micron$. The  blue shading indicates the standard error of the mean.  \label{PACS_WISE}}
\end{figure}

To investigate  the nature of the  16 undetected sources in both PACS bands, we used  values from the {\it WISE} All-sky source catalog\footnote{http://irsa.ipac.caltech.edu/Missions/wise.html} at 3.4, 4.6, 12, and 22 $\micron$ to construct  the AGN spectral energy distribution (SED) between 3.4 and 160~$\micron$ for the BAT sample. From this,  we  compared mean and median  SEDs for the entire sample  with those from the PACS undetected sources.  The upper panel in Figure~\ref{PACS_WISE} shows a comparison between the SEDs of the  sources with detections in both PACS bands and the undetected PACS sources (the  individual SED's of the BAT AGN  are normalized to the flux in the {\it WISE} 22 $\micron$ band). It can be seen that the undetected sources are characterized by a flatter  infrared SED  than that for the entire sample. Note that  the entire sample was detected by {\it WISE}, except for Mrk~3. For the {\it WISE} fluxes, we selected the magnitude measured with profile-fitting photometry for sources flagged by {\it WISE} as point-sources, and for extended sources, we selected the magnitude measured  via elliptical aperture photometry which are measured using areas that are scaled from the 2MASS Extended Source Catalog morphologies \citep{2006AJ....131.1163S}. In addition,  we inspected  the low-resolution MIR   spectra of 6  PACS undetected sources observed by  {\it Spitzer} (lower panel in Figure\ref{PACS_WISE}). We retrieved  the  low-resolution MIR  spectra for these sources, namely, ESO121-IG028, Mrk352, Mrk50, PG2304+042, SBS1301+540 and UM614, via the  Cornell atlas of Spitzer/IRS sources \citep{2011ApJS..196....8L}. From this comparison it is clear that, on average, these sources show a systematic decrease in their   MIR emission towards longer wavelengths, $\lambda > 20 \micron$. Moreover, the low-resolution MIR spectra  show no strong star formation features  (e.g.,  polycyclic aromatic hydrocarbon emission). Overall,  these results suggest  that PACS undetected BAT  AGN reside in  cold  dust depleted systems with no active star formation. We define this group of objects   as X-ray Bright Far-infrared Faint sources (XBFF).   A more detailed analysis of the BAT SEDs and the nature of the XBFF sources will be presented in subsequent papers in the BAT {\it Herschel}  series. 

\begin{figure}
\epsscale{.80}
\plotone{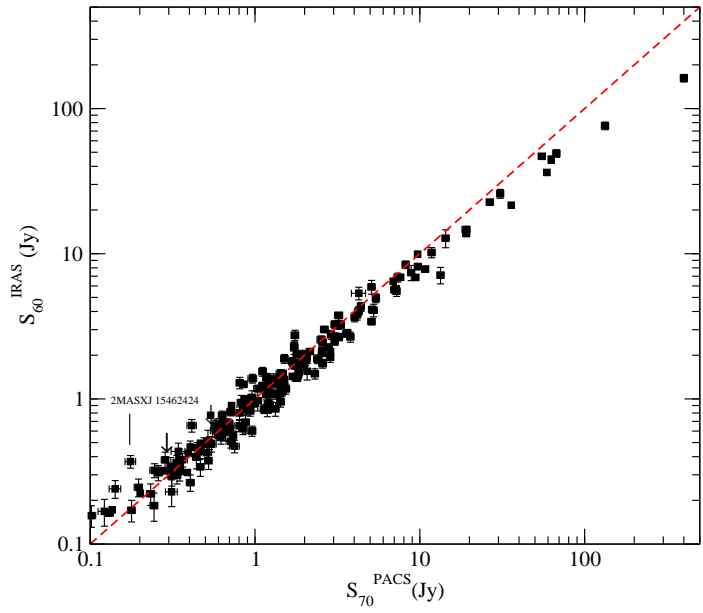}
\caption{Comparison between  the PACS fluxes at 70~$\micron$ (S$_{70}$) and the {\it IRAS} fluxes at 60~$\micron$  (S$_{60}$). It can be seen that the bulk of the {\it IRAS} flux is contained within the PACS (smaller) apertures. The red dashed line indicates $y=x$.  \label{PACS_IRAS}}
\end{figure}

Figure~\ref{PACS_IRAS} shows the comparison between the PACS 70~$\micron$ measurements and observations from the {\it Infrared Astronomical Satellite} at 60~$\micron$ \citep[{\it IRAS},][]{1984ApJ...278L...1N}. Note that in most cases, given PACS better photometric sensitivity, it was not possible to find {\it IRAS} fluxes for all galaxies in the sample. In total, 205 sources from the BAT sample had {\it IRAS} detections at 60 $\micron$, mainly from the {\it IRAS} Faint Source Catalog for faint point sources \citep{1990BAAS...22Q1325M}, the  Point Source Catalog \citep[V2.0, ][]{1988iras....7.....H} and the catalog  of large optical galaxies \citep{1988ApJS...68...91R}.  From this comparison, we found  very good agreement between the PACS and {\it IRAS} fluxes with S$_{60}$ density fluxes ({\it IRAS}) slightly below the 1:1 line, as expected \citep[an increasing flux density at wavelengths shorter than the FIR peak, ][]{2001ApJ...549..215D,2002ApJ...576..159D,2007ApJ...657..810D,2012MNRAS.425.3094C,2012ApJ...745...95D}.  The good correlation between the fluxes suggests that the size of our aperture photometry extractions encompass the FIR emission from the galaxy, because the beam size used for the {\it IRAS} catalog  is  larger, approximately 1\arcmin~ at 60~$\micron$,  than the majority of the aperture sizes used in this work. This result is corroborated by the spatial analysis presented in \cite{2014ApJ...781L..34M}, where we show that in the majority of the BAT sources,  the bulk of the FIR radiation is point-like at the spatial resolution of {\it Herschel} (a median value of 2 kpc FWHM).   Note that some of the brightest sources in our sample, e.g., Centaurus A,  are extended;   thus, the point source extraction from {\it IRAS}, even at 1\arcmin~  resolution, underpredicts the flux at 60~$\micron$.  On the other hand, there may be contamination from a neighboring source that lies  within  {\it IRAS} bigger aperture, therby,  overpredicting the flux at 60~$\micron$. For example, there is an IR bright companion source (GALEXASC J154634.12+692844.7) at $\sim$56.4\arcsec~ from 2MASXJ15462424+6929102 (see Figure~~\ref{PACS_IRAS}).

\begin{deluxetable}{lcccccccccc}
\tabletypesize{\scriptsize}
\tablecolumns{11}
\rotate
\tablecaption{Galaxy Sample and Far-Infrared Flux Densities}
\tablewidth{0pt}
\tablehead{
\colhead{Name} & \colhead{RA} & \colhead{DEC}&
\colhead{Distance} & \colhead{Type} & \colhead{70$\micron$} &
\colhead{160$\micron$} & \colhead{BAT} & \colhead{Aperture} &
\colhead{Aperture} & \colhead{Program}\\
\colhead{} & \colhead{(J2000.0)} & \colhead{(J2000.0)}&
\colhead{Mpc} & \colhead{} & \colhead{Jy} &
\colhead{Jy} & \colhead{BAT} & \colhead{70$\micron$} &
\colhead{160$\micron$} & \colhead{ID} \\
\tableline
\colhead{(1)} & \colhead{(2)} & \colhead{(3)}&
\colhead{(4)} & \colhead{(5)} & \colhead{(6)} &
\colhead{(7)} & \colhead{(8)} & \colhead{(9)} &
\colhead{(10)} & \colhead{(11)} 
}
\startdata

Mrk 335                        	&	00 06 19.5 	&	+20 12 10 	&	111.10	&	Sy 1.2	&	0.309	$\pm$	0.016	&	0.150	$\pm$	0.010	&	18.43	&	12				&	22				&	2   	\\
2MASX J00253292+6821442        	&	00 25 32.9 	&	+68 21 44 	&	51.15	&	Sy 2	&	0.290	$\pm$	0.015	&	0.303	$\pm$	0.024	&	18.24	&	12				&	22				&	1   	\\
CGCG 535-012                   	&	00 36 21.0 	&	+45 39 54 	&	208.70	&	Sy 1.2	&	0.145	$\pm$	0.011	&		$<$	0.177	&	15.58	&	12				&	22				&	1   	\\
NGC 235A                       	&	00 42 52.8 	&	-23 32 28 	&	95.51	&	Sy 1	&	2.337	$\pm$	0.123	&	2.613	$\pm$	0.173	&	47.65	&	12				&	22				&	1   	\\
MCG-02-02-095                 	&	00 43 08.8 	&	-11 36 04 	&	80.87	&	Sy 2	&	0.081	$\pm$	0.008	&		$<$	0.196	&	8.95	&	5.5				&	10.5				&	1   	\\
Mrk 348\tablenotemark{b}                        	&	00 48 47.1 	&	+31 57 25 	&	64.24	&	Sy 2 	&	0.808	$\pm$	0.041	&	0.995	$\pm$	0.050	&	156.04	&	12				&	22				&	3	\\
MCG+05-03-013                 	&	00 51 35.0 	&	+29 24 05 	&	156.12	&	Sy 1	&	0.825	$\pm$	0.046	&	1.890	$\pm$	0.166	&	9.81	&	14				&	24				&	1   	\\
Mrk 352                        	&	00 59 53.3 	&	+31 49 37 	&	63.50	&	Sy 1	&		$<$	0.080	&		$<$	0.089	&	29.98	&	12				&	22				&	1   	\\
ESO 195-IG021NED03             	&	01 00 35.0 	&	-47 52 04 	&	211.44	&	Sy 1.8	&	0.411	$\pm$	0.021	&	0.824	$\pm$	0.106	&	16.30	&	12				&	22				&	1   	\\
MCG-07-03-007                 	&	01 05 26.8 	&	-42 12 58 	&	129.16	&	Sy 2	&	0.313	$\pm$	0.017	&	0.457	$\pm$	0.029	&	11.97	&	12				&	22				&	1   	\\
2MASX J01064523+0638015        	&	01 06 45.3 	&	+06 38 02 	&	178.71	&	Sy 2	&	0.150	$\pm$	0.011	&		$<$	0.097	&	15.45	&	12				&	22				&	1   	\\
2MASX J01073963-1139117        	&	01 07 39.6 	&	-11 39 12 	&	207.86	&	Sy 2	&	0.491	$\pm$	0.033	&	0.715	$\pm$	0.060	&	10.48	&	12				&	22				&	1   	\\
NGC 424                        	&	01 11 27.6 	&	-38 05 00 	&	50.14	&	Sy 2	&	1.687	$\pm$	0.089	&	1.602	$\pm$	0.144	&	20.72	&	12				&	22				&	1   	\\
Mrk 975                        	&	01 13 51.0 	&	+13 16 18 	&	217.74	&	Sy 1	&	0.806	$\pm$	0.050	&	1.279	$\pm$	0.068	&	16.54	&	12				&	22				&	1   	\\
IC 1657                        	&	01 14 07.0 	&	-32 39 03 	&	50.95	&	Sy 2	&	2.895	$\pm$	0.145	&	3.950	$\pm$	0.198	&	14.32	&		80	33	18	&		80	33	18	&	1   	\\
Fairall 9\tablenotemark{b}                      	&	01 23 45.8 	&	-58 48 21 	&	205.85	&	Sy 1	&	0.532	$\pm$	0.039	&	0.611	$\pm$	0.038	&	49.46	&	12				&	18				&	1   	\\
NGC 526A                       	&	01 23 54.4 	&	-35 03 56 	&	81.86	&	Sy 1.5	&	0.201	$\pm$	0.013	&	0.291	$\pm$	0.031	&	63.25	&	12				&	22				&	1   	\\
NGC 513                        	&	01 24 26.9 	&	+33 47 58 	&	83.80	&	Sy 2	&	2.862	$\pm$	0.146	&	4.143	$\pm$	0.211	&	20.35	&	18				&	22				&	1   	\\
Mrk 359                        	&	01 27 32.6 	&	+19 10 44 	&	74.42	&	Sy 1.5	&	1.468	$\pm$	0.078	&	1.473	$\pm$	0.107	&	13.32	&	12				&	22				&	1   	\\
MCG-03-04-072                 	&	01 28 06.7 	&	-18 48 31 	&	201.25	&	Sy 1 	&	0.115	$\pm$	0.009	&	0.225	$\pm$	0.017	&	19.74	&	12				&	22				&	1   	\\
ESO 244-IG030                  	&	01 29 51.2 	&	-42 19 35 	&	110.27	&	Sy 2	&	1.425	$\pm$	0.075	&	2.207	$\pm$	0.116	&	8.86	&		50	22	36	&		50	22	36	&	1   	\\
ESO 297-018\tablenotemark{b}                    	&	01 38 37.2 	&	-40 00 41 	&	108.64	&	Sy 2	&	0.681	$\pm$	0.037	&	1.838	$\pm$	0.094	&	68.66	&	145	17	20		&		145	17	20	&	1   	\\
MCG-01-05-047                 	&	01 52 49.0 	&	-03 26 49 	&	73.60	&	Sy 2	&	2.403	$\pm$	0.135	&	6.131	$\pm$	0.312	&	22.86	&		340	19	63	&		340	19	63	&	1   	\\
UGC 01479                      	&	02 00 19.1 	&	+24 28 25 	&	70.30	&	Sy 2	&	1.748	$\pm$	0.090	&	2.665	$\pm$	0.135	&	14.36	&		0	21	32	&		0	21	32	&	1   	\\
NGC 788                        	&	02 01 06.4 	&	-06 48 56 	&	58.06	&	Sy 2	&	0.518	$\pm$	0.027	&	0.516	$\pm$	0.027	&	80.13	&	12				&	22				&	1   	\\
Mrk 1018                       	&	02 06 16.0 	&	-00 17 29 	&	185.16	&	Sy 1.5	&	0.068	$\pm$	0.012	&	0.115	$\pm$	0.015	&	32.97	&	5.5				&	10.5				&	1   	\\
LEDA 138501                    	&	02 09 34.3 	&	+52 26 33 	&	215.77	&	Sy 1	&		$<$	0.083	&		$<$	0.085	&	51.28	&	12				&	22				&	1   	\\
ESO 197-G027                   	&	02 10 52.5 	&	-49 41 55 	&	210.72	&	Sy 2	&	0.736	$\pm$	0.038	&	1.561	$\pm$	0.085	&	7.88	&	14				&	22				&	1   	\\
Mrk 590                        	&	02 14 33.6 	&	-00 46 00 	&	113.73	&	Sy 1.2	&	0.545	$\pm$	0.036	&	1.877	$\pm$	0.111	&	16.66	&	12				&	22				&	1   	\\
NGC 931                        	&	02 28 14.5 	&	+31 18 42 	&	71.24	&	Sy 1.5	&	2.521	$\pm$	0.143	&	5.331	$\pm$	0.273	&	60.55	&		70	30	67	&		70	30	67	&	1   	\\
IC 1816                        	&	02 31 51.0 	&	-36 40 19 	&	72.51	&	Sy 1.8	&	1.782	$\pm$	0.099	&	2.781	$\pm$	0.154	&	19.25	&	22				&	22				&	1   	\\
NGC 985                        	&	02 34 37.8 	&	-08 47 15 	&	188.35	&	Sy 1	&	1.296	$\pm$	0.070	&	1.866	$\pm$	0.100	&	31.90	&		85	25	30	&		85	25	30	&	1   	\\
ESO 198-024                    	&	02 38 19.7 	&	-52 11 33 	&	198.99	&	Sy 1	&	0.106	$\pm$	0.012	&		$<$	0.110	&	29.56	&	12				&	22				&	1   	\\
NGC 1052                       	&	02 41 04.8 	&	-08 15 21 	&	19.48	&	Sy 2	&	0.855	$\pm$	0.048	&	0.799	$\pm$	0.083	&	29.46	&	12				&	22				&	1   	\\
Mrk 595                        	&	02 41 34.9 	&	+07 11 14 	&	116.36	&	Sy 1.5	&	0.389	$\pm$	0.020	&	0.521	$\pm$	0.033	&	10.86	&	12				&	22				&	1   	\\
ESO 479-G031                   	&	02 44 47.7 	&	-24 30 50 	&	101.19	&	L	&	0.086	$\pm$	0.011	&		$<$	0.152	&	9.05	&	5.5				&	10.5				&	1   	\\
HB 890241+622                  	&	02 44 57.7 	&	+62 28 07 	&	192.21	&	Sy 1	&	0.538	$\pm$	0.027	&	0.362	$\pm$	0.024	&	90.54	&	12				&	22				&	1   	\\
NGC 1106                       	&	02 50 40.5 	&	+41 40 17 	&	61.79	&	Sy 2	&	1.112	$\pm$	0.058	&	1.737	$\pm$	0.123	&	18.58	&	12				&	22				&	1   	\\
2MFGC 02280                    	&	02 50 42.6 	&	+54 42 18 	&	64.76	&	Sy 2	&	1.541	$\pm$	0.081	&	2.063	$\pm$	0.137	&	27.07	&	12				&	22				&	1   	\\
NGC 1125\tablenotemark{b}                       	&	02 51 40.3 	&	-16 39 04 	&	46.56	&	Sy 2	&	3.224	$\pm$	0.164	&	2.376	$\pm$	0.163	&	17.43	&	18				&	24				&	1   	\\
MCG-02-08-014                 	&	02 52 23.4 	&	-08 30 37 	&	71.67	&	Sy 2 	&	0.337	$\pm$	0.019	&	0.541	$\pm$	0.041	&	26.07	&	12				&	22				&	1   	\\
ESO 417-G006                   	&	02 56 21.5 	&	-32 11 08 	&	69.68	&	Sy 2	&	0.244	$\pm$	0.013	&	0.142	$\pm$	0.027	&	30.66	&	12				&	22				&	1   	\\
MCG-02-08-038                 	&	03 00 04.3 	&	-10 49 29 	&	141.14	&	Sy 1	&	0.126	$\pm$	0.010	&	0.340	$\pm$	0.028	&	17.63	&	5.5				&		50	19	30	&	1   	\\
NGC 1194                       	&	03 03 49.1 	&	-01 06 13 	&	58.03	&	Sy 1	&	0.633	$\pm$	0.034	&	0.473	$\pm$	0.025	&	36.58	&	12				&	22				&	1   	\\
ESO 031-G008                   	&	03 07 35.3 	&	-72 50 03 	&	119.17	&	Sy 1.2	&	0.069	$\pm$	0.005	&		$<$	0.259	&	12.43	&	5.5				&	10.5				&	1   	\\
MCG+00-09-042                 	&	03 17 02.2 	&	+01 15 18 	&	102.22	&	Sy 2 / L	&	3.793	$\pm$	0.192	&	4.155	$\pm$	0.241	&	13.27	&	12				&	22				&	1   	\\
IRAS 03219+4031                	&	03 25 13.2 	&	+40 41 55 	&	209.07	&	Sy 2	&	1.306	$\pm$	0.069	&	1.171	$\pm$	0.088	&	19.41	&	14				&	22				&	1   	\\
2MASX J03305218+0538253        	&	03 30 52.2 	&	+05 38 26 	&	201.25	&	Sy 1	&	0.222	$\pm$	0.016	&	0.130	$\pm$	0.020	&	11.02	&	5.5				&	10.5				&	1   	\\
MCG-01-09-045                 	&	03 31 23.0 	&	-05 08 30 	&	54.69	&	Sy 2	&		$<$	0.133	&	0.367	$\pm$	0.025	&	12.75	&	12				&	22				&	1   	\\
NGC 1365                       	&	03 33 36.4 	&	-36 08 25 	&	17.93	&	Sy 1.8	&	133.097	$\pm$	6.656	&	217.359	$\pm$	10.871	&	64.07	&		30	139	213	&		30	185	336	&	3	\\
2MASX J03342453-1513402\tablenotemark{b}        	&	03 34 24.5 	&	-15 13 40 	&	151.59	&	Sy 1.5	&	0.610	$\pm$	0.039	&	0.974	$\pm$	0.079	&	12.46	&	12				&	22				&	1   	\\
ESO 548-G081                   	&	03 42 03.7 	&	-21 14 39 	&	61.84	&	Sy 1	&	0.960	$\pm$	0.051	&	2.181	$\pm$	0.132	&	44.46	&	18				&	22				&	1   	\\
LCRS B034324.7-394349\tablenotemark{b}          	&	03 45 12.5 	&	-39 34 29 	&	186.73	&	Sy 1	&	0.142	$\pm$	0.011	&	0.224	$\pm$	0.023	&	10.36	&	5.5				&	10.5				&	1   	\\
2MASX J03502377-5018354\tablenotemark{b}        	&	03 50 23.8 	&	-50 18 36 	&	158.52	&	Sy 2	&	0.395	$\pm$	0.027	&	0.471	$\pm$	0.032	&	20.89	&	12				&	10.5				&	1   	\\
2MASX J03534246+3714077        	&	03 53 42.5 	&	+37 14 07 	&	79.90	&	Sy 2	&	0.817	$\pm$	0.046	&	1.128	$\pm$	0.087	&	15.33	&	12				&	22				&	1   	\\
2MASX J03540948+0249307        	&	03 54 09.5 	&	+02 49 31 	&	156.32	&	Sy 1	&	0.183	$\pm$	0.012	&		$<$	0.198	&	12.85	&	5.5				&	10.5				&	1   	\\
ESO 549-G049                   	&	04 02 25.7 	&	-18 02 51 	&	113.32	&	Sy 2 / L	&	3.629	$\pm$	0.188	&	5.131	$\pm$	0.262	&	25.16	&	18				&	22				&	1   	\\
IRAS 04124-0803                	&	04 14 52.7 	&	-07 55 40 	&	165.97	&	Sy 1	&	0.657	$\pm$	0.049	&	0.439	$\pm$	0.024	&	21.00	&	12				&	22				&	1   	\\
3C 111                       	&	04 18 21.3 	&	+38 01 36 	&	212.59	&	Sy 1	&	0.266	$\pm$	0.017	&	0.532	$\pm$	0.062	&	116.80	&	12				&	22				&	1   	\\
ESO 157-G023\tablenotemark{b}                   	&	04 22 24.2 	&	-56 13 33 	&	190.09	&	Sy 2	&	0.082	$\pm$	0.007	&	0.588	$\pm$	0.034	&	21.20	&	5.5				&		150	17	28	&	1   	\\
2MASX J04234080+0408017\tablenotemark{b}        	&	04 23 40.8 	&	+04 08 02 	&	196.73	&	Sy 2	&	0.574	$\pm$	0.038	&	0.502	$\pm$	0.028	&	23.88	&	12				&	10.5				&	1   	\\
3C 120                         	&	04 33 11.1 	&	+05 21 16 	&	143.01	&	Sy 1	&	1.368	$\pm$	0.070	&	1.639	$\pm$	0.102	&	94.36	&	12				&	22				&	1   	\\
Mrk 618                        	&	04 36 22.2 	&	-10 22 34 	&	154.31	&	Sy 1	&	3.036	$\pm$	0.166	&	3.244	$\pm$	0.195	&	17.75	&	18				&	28				&	1   	\\
MCG-02-12-050\tablenotemark{b}                 	&	04 38 14.2 	&	-10 47 45 	&	157.89	&	Sy 1.2	&	0.603	$\pm$	0.042	&	1.376	$\pm$	0.076	&	18.97	&	40	19	23		&		40	19	23	&	1   	\\
1RXS J044154.5-082639          	&	04 41 54.1 	&	-08 26 34 	&	192.21	&	Sy 1	&	0.176	$\pm$	0.012	&	0.149	$\pm$	0.024	&	10.21	&	12				&	22				&	1   	\\
UGC 03142                      	&	04 43 46.8 	&	+28 58 19 	&	93.00	&	Sy 1	&	1.227	$\pm$	0.068	&	2.659	$\pm$	0.161	&	45.54	&	35				&	35				&	1   	\\
2MASX J04440903+2813003        	&	04 44 09.0 	&	+28 13 01 	&	48.00	&	Sy 2	&	1.224	$\pm$	0.069	&	2.411	$\pm$	0.340	&	52.95	&	12				&	22				&	1   	\\
MCG-01-13-025                 	&	04 51 41.5 	&	-03 48 33 	&	67.96	&	Sy 1.2	&	0.092	$\pm$	0.008	&	0.160	$\pm$	0.030	&	31.11	&	5.5				&	10.5				&	1   	\\
1RXS J045205.0+493248          	&	04 52 05.0 	&	+49 32 45 	&	125.26	&	Sy 1	&	0.354	$\pm$	0.021	&	0.785	$\pm$	0.086	&	63.16	&	12				&	22				&	1   	\\
CGCG 420-015                   	&	04 53 25.8 	&	+04 03 42 	&	126.98	&	Sy 2	&	0.648	$\pm$	0.034	&	0.684	$\pm$	0.061	&	28.14	&	12				&	22				&	1   	\\
ESO 033-G002                   	&	04 55 59.0 	&	-75 32 28 	&	77.52	&	Sy 2	&	0.697	$\pm$	0.038	&	0.689	$\pm$	0.083	&	21.26	&	12				&	22				&	1   	\\
2MASX J05020903+0331499        	&	05 02 09.0 	&	+03 31 50 	&	68.37	&	Sy 1	&	0.084	$\pm$	0.010	&		$<$	0.183	&	15.31	&	12				&	22				&	1   	\\
2MASX J05054575-2351139        	&	05 05 45.7 	&	-23 51 14 	&	152.05	&	Sy 2 / HII	&	0.179	$\pm$	0.011	&	0.142	$\pm$	0.025	&	60.85	&	12				&	22				&	1   	\\
CGCG 468-002NED01\tablenotemark{b}              	&	05 08 19.7 	&	+17 21 48 	&	74.94	&	Sy 2	&	2.323	$\pm$	0.119	&	2.378	$\pm$	0.143	&	26.03	&	12				&	12				&	1   	\\
IRAS 05078+1626                	&	05 10 45.5 	&	+16 29 56 	&	76.56	&	Sy 1.5	&	1.175	$\pm$	0.061	&	0.727	$\pm$	0.090	&	90.61	&	12				&	22				&	1   	\\
ESO 553-G022                   	&	05 11 57.8 	&	-18 29 38 	&	183.56	&	Sy 2	&	0.082	$\pm$	0.010	&	0.309	$\pm$	0.036	&	13.72	&	12				&	22				&	1   	\\
ARK 120                        	&	05 16 11.4 	&	-00 08 59 	&	141.69	&	Sy 1	&	0.685	$\pm$	0.039	&	1.176	$\pm$	0.097	&	69.95	&	12				&	22				&	1   	\\
MCG-02-14-009                 	&	05 16 21.2 	&	-10 33 41 	&	122.84	&	Sy 1	&	0.343	$\pm$	0.033	&	0.813	$\pm$	0.060	&	13.58	&	12				&	22				&	1   	\\
ESO 362-18                     	&	05 19 35.8 	&	-32 39 28 	&	53.07	&	Sy 1.5	&	1.426	$\pm$	0.073	&	1.923	$\pm$	0.098	&	49.31	&	25				&	25				&	1   	\\
PICTOR A                       	&	05 19 49.7 	&	-45 46 44 	&	152.12	&	Sy 1 / L	&	0.130	$\pm$	0.011	&	0.305	$\pm$	0.023	&	38.59	&	12				&	22				&	1   	\\
2MASX J05240693-1210087\tablenotemark{b}	&	05 24 06.5 	&	-12 10 00 	&	214.86	&	Sy 1	&	0.330	$\pm$	0.021	&	0.261	$\pm$	0.021	&	18.34	&	5.5				&	10.5				&	1   	\\
ESO 553-G043                   	&	05 26 27.3 	&	-21 17 12 	&	119.77	&	Sy 2	&	0.122	$\pm$	0.010	&		$<$	0.114	&	13.85	&	12				&	22				&	1   	\\
NGC 2110                       	&	05 52 11.4 	&	-07 27 22 	&	35.60	&	Sy 2	&	5.141	$\pm$	0.261	&	5.227	$\pm$	0.267	&	319.13	&	18				&	22				&	1   	\\
MCG+08-11-011                 	&	05 54 53.6 	&	+46 26 22 	&	87.90	&	Sy 1.5	&	2.639	$\pm$	0.142	&	3.467	$\pm$	0.178	&	133.44	&		0	25	54	&		0	34	58	&	1   	\\
2MASX J05580206-3820043        	&	05 58 02.0 	&	-38 20 05 	&	146.83	&	Sy 1	&	0.246	$\pm$	0.016	&		$<$	0.199	&	29.27	&	5.5				&	10.5				&	1   	\\
IRAS 05589+2828                	&	06 02 10.7 	&	+28 28 22 	&	142.97	&	Sy 1	&	0.959	$\pm$	0.055	&	0.813	$\pm$	0.094	&	65.19	&	12				&	22				&	1   	\\
ESO 005-G004                   	&	06 05 41.6 	&	-86 37 55 	&	22.40	&	Sy 2	&	7.616	$\pm$	0.387	&	22.206	$\pm$	1.111	&	33.60	&		95	11	56	&		95	26	79	&	1   	\\
Mrk 3	&	06 15 36.3 	&	71 02 15 	&	57.70	&	Sy 2	&	3.220	$\pm$	0.183	&	2.530	$\pm$	0.166	&	135.71	&	38				&	38				&	7	\\
ESO 121-IG028\tablenotemark{b}                  	&	06 23 45.6 	&	-60 58 44 	&	176.55	&	Sy 2	&		$<$	0.075	&		$<$	0.233	&	28.12	&	12				&	22				&	1   	\\
ESO 426-G002                   	&	06 23 46.4 	&	-32 13 00 	&	96.39	&	Sy 2	&	0.196	$\pm$	0.011	&	0.146	$\pm$	0.022	&	24.36	&	5.5				&	10.5				&	1   	\\
6dF J0626586-370559            	&	06 26 58.6 	&	-37 05 59 	&	167.57	&	Sy 1.5	&	0.209	$\pm$	0.013	&	0.442	$\pm$	0.032	&	11.69	&	12				&	22				&	1   	\\
VII Zw 073                      	&	06 30 25.6 	&	+63 40 41 	&	180.20	&	Sy 2	&	2.056	$\pm$	0.104	&	1.804	$\pm$	0.138	&	11.14	&	12				&	22				&	1   	\\
UGC 03478                      	&	06 32 47.2 	&	+63 40 25 	&	54.46	&	Sy 1.2	&	1.465	$\pm$	0.077	&	3.824	$\pm$	0.197	&	10.13	&		50	27	41	&		50	27	41	&	1   	\\
ESO 490-IG026                  	&	06 40 11.7 	&	-25 53 43 	&	106.99	&	Sy 1.2	&	1.975	$\pm$	0.102	&	1.604	$\pm$	0.103	&	37.42	&	12				&	22				&	1   	\\
2MASX J06411806+3249313        	&	06 41 18.0 	&	+32 49 32 	&	205.78	&	Sy 2	&	0.087	$\pm$	0.009	&		$<$	0.053	&	35.08	&	12				&	22				&	1   	\\
Mrk 6\tablenotemark{b}                          	&	06 52 12.3 	&	+74 25 37 	&	80.62	&	Sy 1.5	&	1.043	$\pm$	0.061	&	0.942	$\pm$	0.071	&	60.60	&	12				&	18				&	1   	\\
UGC 03601                      	&	06 55 49.5 	&	+40 00 01 	&	73.28	&	Sy 1.5	&	0.367	$\pm$	0.020	&	0.719	$\pm$	0.075	&	21.38	&	12				&	22				&	1   	\\
2MASX J06561197-4919499        	&	06 56 12.0 	&	-49 19 50 	&	178.71	&	Sy 2	&	0.259	$\pm$	0.016	&	0.126	$\pm$	0.023	&	12.17	&	5.5				&	10.5				&	1   	\\
MCG+06-16-028                 	&	07 14 03.9 	&	+35 16 45 	&	67.09	&	Sy 2	&	2.549	$\pm$	0.128	&	2.487	$\pm$	0.155	&	16.71	&	12				&	22				&	1   	\\
2MASX J07262635-3554214        	&	07 26 26.4 	&	-35 54 22 	&	127.04	&	Sy 2	&	1.113	$\pm$	0.059	&		$<$	0.700	&	23.35	&	12				&	22\tablenotemark{a}				&	1   	\\
Mrk 79                         	&	07 42 32.8 	&	+49 48 35 	&	95.34	&	Sy 1.2	&	1.431	$\pm$	0.088	&	2.370	$\pm$	0.185	&	46.57	&	22				&	30				&	1   	\\
UGC 03995A\tablenotemark{b}                     	&	07 44 07.0 	&	+29 14 57 	&	67.60	&	Sy 2 	&		$<$	0.093	&	1.915	$\pm$	0.097	&	16.98	&	12				&	22				&	1   	\\
Mrk 10                         	&	07 47 29.1 	&	+60 56 01 	&	126.38	&	Sy 1.2	&	0.713	$\pm$	0.040	&	1.973	$\pm$	0.101	&	14.80	&		130	21	31	&		130	26	37	&	1   	\\
2MASS J07594181-3843560        	&	07 59 41.7 	&	-38 43 57 	&	174.22	&	Sy 1.2	&	0.046	$\pm$	0.007	&		$<$	0.121	&	50.64	&	5.5				&	10.5				&	1   	\\
2MASX J07595347+2323241        	&	07 59 53.5 	&	+23 23 24 	&	126.06	&	Sy 2	&	2.311	$\pm$	0.120	&	4.943	$\pm$	0.250	&	34.54	&	18				&	22				&	1   	\\
IC 0486                        	&	08 00 21.0 	&	+26 36 49 	&	115.89	&	Sy 1	&	0.989	$\pm$	0.062	&	1.483	$\pm$	0.089	&	32.16	&	14				&	22				&	1   	\\
ESO 209-G012                   	&	08 01 58.0 	&	-49 46 42 	&	176.44	&	Sy 1.5	&	1.039	$\pm$	0.058	&	2.077	$\pm$	0.107	&	21.27	&	28				&	28				&	1   	\\
2MASX J08032736+0841523        	&	08 03 27.4 	&	+08 41 52 	&	204.87	&	Sy 1	&		$<$	0.079	&		$<$	0.112	&	17.32	&	12				&	22				&	1   	\\
Mrk 1210                       	&	08 04 05.9 	&	+05 06 50 	&	57.60	&	Sy 2	&	1.510	$\pm$	0.078	&	0.914	$\pm$	0.089	&	54.44	&	12				&	22				&	1   	\\
MCG+02-21-013                 	&	08 04 46.4 	&	+10 46 36 	&	149.34	&	Sy 2	&	0.713	$\pm$	0.046	&	1.701	$\pm$	0.103	&	15.22	&	12				&	22				&	1   	\\
Fairall 272\tablenotemark{b}                    	&	08 23 01.1 	&	-04 56 05 	&	93.70	&	Sy 2	&	0.798	$\pm$	0.044	&	1.222	$\pm$	0.071	&	46.43	&	12				&	22				&	1   	\\
Fairall 1146                   	&	08 38 30.8 	&	-35 59 33 	&	136.66	&	Sy 1.5	&	1.219	$\pm$	0.068	&	1.407	$\pm$	0.090	&	28.91	&	14				&	22				&	1   	\\
MCG+11-11-032                 	&	08 55 12.5 	&	+64 23 46 	&	157.45	&	Sy 2	&	0.113	$\pm$	0.008	&	0.317	$\pm$	0.020	&	17.71	&	12				&	22				&	1   	\\
NGC 2655                       	&	08 55 37.7 	&	+78 13 23 	&	24.40	&	Sy 2	&	1.830	$\pm$	0.093	&	2.862	$\pm$	0.147	&	13.44	&	14				&	24				&	1   	\\
Mrk 18                         	&	09 01 58.4 	&	+60 09 06 	&	47.23	&	Sy 2	&	2.565	$\pm$	0.133	&	2.409	$\pm$	0.170	&	12.21	&	12				&	22				&	1   	\\
2MASX J09023729-4813339        	&	09 02 37.3 	&	-48 13 34 	&	170.18	&	Sy 1	&	0.092	$\pm$	0.010	&		$<$	0.801	&	29.41	&	12				&	22\tablenotemark{a}				&	1   	\\
2MASX J09043699+5536025        	&	09 04 36.9 	&	+55 36 03 	&	161.42	&	Sy 1	&	0.181	$\pm$	0.010	&		$<$	0.130	&	15.20	&	12				&	22				&	1   	\\
Mrk 704                        	&	09 18 26.0 	&	+16 18 19 	&	126.29	&	Sy 1.5	&	0.336	$\pm$	0.018	&	0.253	$\pm$	0.027	&	33.09	&	12				&	22				&	1   	\\
SBS 0915+556\tablenotemark{b}                   	&	09 19 13.2 	&	+55 27 55 	&	216.68	&	Sy 2	&	0.098	$\pm$	0.013	&		$<$	0.075	&	8.13	&	5.5				&	10.5				&	1   	\\
IC 2461                        	&	09 19 58.0 	&	+37 11 29 	&	54.18	&	Sy 1.9	&	1.446	$\pm$	0.075	&	3.244	$\pm$	0.164	&	19.91	&		48	40	15	&		48	40	15	&	1   	\\
MCG-01-24-012\tablenotemark{b}                 	&	09 20 46.2 	&	-08 03 22 	&	84.24	&	Sy 2	&	0.572	$\pm$	0.031	&	1.091	$\pm$	0.136	&	40.82	&	5.5				&	34				&	1   	\\
MCG+04-22-042                 	&	09 23 43.0 	&	+22 54 33 	&	140.08	&	Sy 1.2	&	0.131	$\pm$	0.012	&	0.399	$\pm$	0.027	&	39.86	&	5.5				&		0	16	28	&	1   	\\
2MASX J09235371-3141305        	&	09 23 53.7 	&	-31 41 31 	&	184.85	&	Sy 1.9	&	0.110	$\pm$	0.007	&		$<$	0.163	&	19.16	&	12				&	22				&	1   	\\
2MASX J09254750+6927532        	&	09 25 47.5 	&	+69 27 53 	&	169.73	&	Sy 1	&	0.086	$\pm$	0.009	&		$<$	0.200	&	9.23	&	5.5				&	10.5				&	1   	\\
NGC 2885                       	&	09 27 18.5 	&	+23 01 12 	&	113.43	&	Sy 1	&	0.233	$\pm$	0.013	&	0.648	$\pm$	0.057	&	15.91	&	12				&	22				&	1   	\\
CGCG 312-012                   	&	09 29 37.9 	&	+62 32 39 	&	110.33	&	Sy 2	&	0.071	$\pm$	0.006	&	0.189	$\pm$	0.027	&	9.04	&	12				&	22				&	1   	\\
ESO 565-G019\tablenotemark{b}                   	&	09 34 43.6 	&	-21 55 40 	&	69.65	&	Sy 2	&	3.045	$\pm$	0.155	&	3.029	$\pm$	0.154	&	20.68	&	18				&	18				&	1   	\\
2MASX J09360622-6548336\tablenotemark{b}        	&	09 36 06.3 	&	-65 48 33 	&	170.72	&	Sy 1.8	&		$<$	0.074	&		$<$	0.118	&	12.97	&	12				&	22				&	1   	\\
CGCG 122-055                   	&	09 42 04.8 	&	+23 41 07 	&	91.76	&	Sy 1	&	0.651	$\pm$	0.040	&	0.572	$\pm$	0.039	&	13.76	&	12				&	22				&	1   	\\
NGC 2992\tablenotemark{b}                       	&	09 45 42.1 	&	-14 19 35 	&	31.60	&	Sy 2	&	9.456	$\pm$	0.480	&	12.119	$\pm$	0.755	&	26.89	&	35				&	60				&	3	\\
MCG-05-23-016                 	&	09 47 40.2 	&	-30 56 55 	&	36.10	&	Sy 2	&	1.454	$\pm$	0.074	&	0.477	$\pm$	0.031	&	201.30	&	12				&	22				&	1   	\\
NGC 3035                       	&	09 51 55.0 	&	-06 49 23 	&	62.03	&	Sy 1.8	&	0.803	$\pm$	0.049	&	2.633	$\pm$	0.134	&	19.32	&		144	31	42	&		144	31	42	&	1   	\\
NGC 3081                       	&	09 59 29.5 	&	-22 49 35 	&	32.50	&	Sy 2	&	2.657	$\pm$	0.134	&	3.992	$\pm$	0.202	&	83.17	&		80	35	49	&		80	35	49	&	3	\\
2MASX J09594263-3112581        	&	09 59 42.6 	&	-31 12 58 	&	160.78	&	Sy 1	&	0.282	$\pm$	0.018	&	0.270	$\pm$	0.027	&	17.24	&	12				&	22				&	1   	\\
NGC 3079 	&	10 01 57.8 	&	+55 40 47 	&	19.13	&	Sy 2	&	62.729	$\pm$	2.155	&	98.058	$\pm$	4.778	&	33.03	&		170	34	113	&		170	34	113	&	1   	\\
ESO 499-G041                   	&	10 05 55.4 	&	-23 03 25 	&	54.67	&	Sy 1.5	&	0.540	$\pm$	0.028	&	0.940	$\pm$	0.137	&	14.63	&	12				&	22				&	1   	\\
ESO 263-G013                   	&	10 09 48.2 	&	-42 48 40 	&	145.35	&	Sy 2	&	0.165	$\pm$	0.011	&		$<$	0.196	&	35.35	&	5.5				&	10.5				&	1   	\\
ESO 374-G044                   	&	10 13 19.9 	&	-35 58 58 	&	122.84	&	Sy 2	&	0.285	$\pm$	0.015	&	0.326	$\pm$	0.032	&	20.07	&	12				&	22				&	1   	\\
ARK 241                        	&	10 21 40.2 	&	-03 27 14 	&	178.01	&	Sy 1	&	0.081	$\pm$	0.005	&		$<$	0.178	&	20.41	&	12				&	22				&	1   	\\
NGC 3227                       	&	10 23 30.6 	&	+19 51 54 	&	20.85	&	Sy 1.5	&	10.781	$\pm$	0.544	&	21.675	$\pm$	1.091	&	109.94	&		152	64	87	&		152	64	87	&	5	\\
NGC 3281                       	&	10 31 52.1 	&	-34 51 13 	&	45.45	&	Sy 2	&	7.342	$\pm$	0.369	&	7.006	$\pm$	0.361	&	86.41	&	24				&	45				&	3	\\
2MASX J10402231-4625264        	&	10 40 22.5 	&	-46 25 26 	&	102.93	&	Sy 2	&	1.202	$\pm$	0.064	&	1.407	$\pm$	0.105	&	24.93	&	12				&	22				&	1   	\\
SDSS J104326.47+110524.2       	&	10 43 26.5 	&	+11 05 24 	&	208.41	&	Sy 1	&	0.058	$\pm$	0.010	&		$<$	0.140	&	14.17	&	12				&	22				&	1   	\\
MCG+12-10-067                 	&	10 44 08.5 	&	+70 24 19 	&	145.53	&	Sy 2	&	0.869	$\pm$	0.046	&	1.908	$\pm$	0.099	&	12.88	&		15	19	32	&		15	19	32	&	1   	\\
MCG+06-24-008                 	&	10 44 49.0 	&	+38 10 52 	&	111.52	&	Sy 1.9	&	0.752	$\pm$	0.051	&	1.850	$\pm$	0.104	&	13.98	&	14				&	22				&	1   	\\
UGC 05881                      	&	10 46 42.5 	&	+25 55 54 	&	88.36	&	Sy 2	&	1.891	$\pm$	0.096	&	2.002	$\pm$	0.128	&	19.31	&	12				&	22				&	1   	\\
NGC 3393                       	&	10 48 23.5 	&	-25 09 43 	&	53.34	&	Sy 2	&	1.731	$\pm$	0.088	&	3.780	$\pm$	0.196	&	25.98	&	18				&	55				&	1   	\\
Mrk 417                        	&	10 49 30.9 	&	+22 57 52 	&	141.88	&	Sy 2	&	0.132	$\pm$	0.008	&	0.118	$\pm$	0.020	&	32.40	&	5.5				&	10.5				&	1   	\\
NGC 3431                       	&	10 51 15.0 	&	-17 00 29 	&	75.01	&	Sy 2	&	0.665	$\pm$	0.051	&	1.745	$\pm$	0.098	&	22.43	&		130	17	39	&		130	17	39	&	1   	\\
Mrk 728                        	&	11 01 01.8 	&	+11 02 49 	&	154.74	&	Sy 1.9	&	0.046	$\pm$	0.005	&		$<$	0.062	&	13.72	&	5.5				&	10.5				&	1   	\\
NGC 3516                       	&	11 06 47.5 	&	+72 34 07 	&	38.90	&	Sy 1.5	&	1.804	$\pm$	0.092	&	1.196	$\pm$	0.090	&	118.34	&	12				&	22				&	1   	\\
IC 2637                        	&	11 13 49.7 	&	+09 35 11 	&	126.27	&	Sy 1.5	&	2.576	$\pm$	0.136	&	4.187	$\pm$	0.221	&	12.39	&	18				&	22				&	1   	\\
ARP 151                        	&	11 25 36.2 	&	+54 22 57 	&	90.54	&	Sy 1	&		$<$	0.048	&		$<$	0.070	&	19.57	&	5.5				&	10.5				&	1   	\\
ESO 439-G009\tablenotemark{b}                   	&	11 27 23.4 	&	-29 15 27 	&	102.78	&	Sy 2 	&	0.600	$\pm$	0.044	&	1.486	$\pm$	0.078	&	6.71	&	102	19	24		&		102	19	24	&	1   	\\
NGC 3718                       	&	11 32 34.9 	&	+53 04 05 	&	17.00	&	L	&	0.632	$\pm$	0.051	&	3.709	$\pm$	0.240	&	12.09	&	14				&		130	23	100	&	1   	\\
IC 2921                        	&	11 32 49.3 	&	+10 17 47 	&	192.38	&	Sy 1	&	0.090	$\pm$	0.011	&		$<$	0.172	&	15.64	&	5.5				&	10.5				&	1   	\\
Mrk 739E                       	&	11 36 29.4 	&	+21 35 46 	&	128.42	&	Sy 1	&	1.683	$\pm$	0.085	&	3.130	$\pm$	0.179	&	13.19	&	14				&	22				&	1   	\\
IGR J11366-6002                	&	11 36 42.0 	&	-60 03 07 	&	59.77	&	Sy 2 / L	&	1.142	$\pm$	0.059	&	1.587	$\pm$	0.090	&	20.69	&	12				&	22				&	1   	\\
NGC 3783                       	&	11 39 01.8 	&	-37 44 19 	&	38.50	&	Sy 1	&	3.046	$\pm$	0.157	&	4.491	$\pm$	0.270	&	181.11	&	40				&	40				&	3	\\
NGC 3786\tablenotemark{b}                       	&	11 39 42.6 	&	+31 54 33 	&	46.18	&	Sy 1.8	&	2.190	$\pm$	0.205	&	3.138	$\pm$	0.454	&	18.02	&	60	19	29		&		60	19	29	&	1   	\\
KUG 1141+371                   	&	11 44 29.9 	&	+36 53 09 	&	165.52	&	Sy 1	&	0.045	$\pm$	0.009	&		$<$	0.098	&	15.38	&	5.5				&	10.5				&	1   	\\
UGC 06728                      	&	11 45 16.0 	&	+79 40 53 	&	27.70	&	Sy 1.2	&	0.104	$\pm$	0.008	&		$<$	0.087	&	27.20	&	12				&	22				&	1   	\\
2MASX J11454045-1827149        	&	11 45 40.5 	&	-18 27 16 	&	142.74	&	Sy 1	&	0.312	$\pm$	0.026	&	0.339	$\pm$	0.024	&	53.26	&	12				&	22				&	1   	\\
MCG+05-28-032\tablenotemark{b}                 	&	11 48 45.9 	&	+29 38 28 	&	98.95	&	L	&	0.876	$\pm$	0.049	&	1.381	$\pm$	0.121	&	24.48	&	12				&	22				&	1   	\\
MCG-01-30-041                 	&	11 52 38.2 	&	-05 12 26 	&	80.63	&	Sy 1.8	&	1.497	$\pm$	0.077	&	1.751	$\pm$	0.124	&	14.02	&	12				&	22				&	1   	\\
2MASX J12005792+0648226        	&	12 00 57.9 	&	+06 48 23 	&	156.52	&	Sy 2	&	0.522	$\pm$	0.027	&	0.917	$\pm$	0.060	&	21.18	&	12				&	22				&	1   	\\
Mrk 1310                       	&	12 01 14.4 	&	-03 40 41 	&	83.87	&	Sy 1	&	0.137	$\pm$	0.011	&	0.268	$\pm$	0.026	&	11.32	&	12				&	22				&	1   	\\
LEDA 38038                     	&	12 02 47.6 	&	-53 50 08 	&	120.69	&	Sy 2	&	1.951	$\pm$	0.102	&	1.413	$\pm$	0.126	&	48.02	&	12				&	22				&	1   	\\
NGC 4051                       	&	12 03 09.6 	&	+44 31 53 	&	14.58	&	Sy 1.5	&	13.365	$\pm$	0.675	&	38.655	$\pm$	1.943	&	39.54	&		120	104	133	&		120	104	133	&	5	\\
ARK 347                        	&	12 04 29.7 	&	+20 18 58 	&	96.46	&	Sy 2	&	0.365	$\pm$	0.019	&	0.497	$\pm$	0.030	&	29.48	&	12				&	22				&	1   	\\
UGC 07064\tablenotemark{b}                      	&	12 04 43.3 	&	+31 10 38 	&	107.64	&	Sy 1.9	&	1.749	$\pm$	0.092	&	3.089	$\pm$	0.175	&	13.27	&	14				&	22				&	1   	\\
NGC 4102                       	&	12 06 23.0 	&	+52 42 40 	&	20.38	&	L	&	55.005	$\pm$	2.751	&	54.194	$\pm$	2.710	&	27.90	&		35	45	55	&		35	45	55	&	1   	\\
Mrk 198                        	&	12 09 14.1 	&	+47 03 30 	&	104.23	&	Sy 2	&	0.846	$\pm$	0.050	&	1.025	$\pm$	0.065	&	22.22	&	12				&	22				&	1   	\\
NGC 4138                       	&	12 09 29.8 	&	+43 41 07 	&	15.57	&	Sy 1.9	&	2.063	$\pm$	0.129	&	5.537	$\pm$	0.459	&	30.00	&		147	32	45	&		147	32	45	&	1   	\\
NGC 4151\tablenotemark{b}                       	&	12 10 32.6 	&	+39 24 21 	&	9.90	&	Sy 1.5	&	6.889	$\pm$	0.354	&	9.055	$\pm$	0.460	&	538.93	&	135	64	96		&		135	64	96	&	4	\\
KUG 1208+386                   	&	12 10 44.3 	&	+38 20 10 	&	97.97	&	Sy 1	&	0.195	$\pm$	0.014	&		$<$	0.163	&	20.69	&	5.5				&	10.5				&	1   	\\
NGC 4180                       	&	12 13 03.0 	&	+07 02 20 	&	40.08	&	Sy 2	&	5.250	$\pm$	0.264	&	8.980	$\pm$	0.451	&	16.66	&		40	29	39	&		40	29	39	&	1   	\\
NGC 4235                       	&	12 17 09.9 	&	+07 11 30 	&	24.97	&	Sy 1	&	0.349	$\pm$	0.018	&	0.823	$\pm$	0.130	&	32.45	&	12				&	22				&	1   	\\
Mrk 202                        	&	12 17 55.0 	&	+58 39 35 	&	90.21	&	Sy 1 	&	0.197	$\pm$	0.012	&	0.295	$\pm$	0.024	&	8.07	&	12				&	22				&	1   	\\
Mrk 766                        	&	12 18 26.5 	&	+29 48 46 	&	55.15	&	Sy 1.5	&	4.259	$\pm$	0.216	&	2.941	$\pm$	0.192	&	21.67	&	12				&	22				&	1   	\\
MESSIER 106                          	&	12 18 57.5 	&	+47 18 14 	&	7.44	&	Sy 1.9 / L	&	35.872	$\pm$	0.761	&	105.177	$\pm$	2.314	&	22.00	&		160	158	376	&		160	158	376	&	3	\\
Mrk 50                         	&	12 23 24.1 	&	+02 40 45 	&	100.78	&	Sy 1	&		$<$	0.055	&		$<$	0.142	&	22.62	&	12				&	22				&	1   	\\
NGC 4388                       	&	12 25 46.7 	&	+12 39 44 	&	21.37	&	Sy 2	&	11.796	$\pm$	0.594	&	18.588	$\pm$	0.933	&	277.38	&		92	44	78	&		92	44	78	&	5	\\
2MASX J12313717-4758019        	&	12 31 37.2 	&	-47 58 02 	&	119.31	&	Sy 1	&	1.507	$\pm$	0.079	&	1.955	$\pm$	0.119	&	15.52	&	12				&	22				&	1   	\\
2MASX J12335145-2103448        	&	12 33 51.4 	&	-21 03 45 	&	99.04	&	Sy 1	&	0.349	$\pm$	0.018	&	0.286	$\pm$	0.028	&	9.89	&	12				&	22				&	1   	\\
NGC 4507                       	&	12 35 36.6 	&	-39 54 33 	&	50.30	&	Sy 2	&	4.392	$\pm$	0.221	&	4.677	$\pm$	0.241	&	188.11	&	35				&	35				&	1   	\\
ESO 506-G027                   	&	12 38 54.6 	&	-27 18 28 	&	107.75	&	Sy 2	&	0.529	$\pm$	0.036	&	1.219	$\pm$	0.066	&	93.94	&	12				&		76	17	37	&	1   	\\
LEDA 170194\tablenotemark{b}                    	&	12 39 06.3 	&	-16 10 47 	&	159.33	&	Sy 2	&	0.373	$\pm$	0.022	&	0.638	$\pm$	0.046	&	41.15	&	12				&	22				&	1   	\\
Mrk 653                        	&	12 39 51.7 	&	+34 58 30 	&	188.01	&	Sy 2	&	0.136	$\pm$	0.009	&	0.412	$\pm$	0.025	&	9.00	&	12				&	22				&	1   	\\
WKK 1263                       	&	12 41 25.7 	&	-57 50 04 	&	105.15	&	Sy 1.5	&	0.728	$\pm$	0.040	&	1.005	$\pm$	0.089	&	46.66	&	12				&	22				&	1   	\\
NGC 4619                       	&	12 41 44.5 	&	+35 03 46 	&	99.35	&	Sy 1	&	1.339	$\pm$	0.072	&	3.236	$\pm$	0.165	&	7.02	&		0	27	33	&		0	27	33	&	1   	\\
2MASX J12475784-5829599\tablenotemark{b}        	&	12 47 57.8 	&	-58 30 00 	&	120.84	&	Sy 1.9	&		$<$	0.093	&		$<$	0.081	&	9.56	&	12				&	22				&	1   	\\
NGC 4748                       	&	12 52 12.5 	&	-13 24 53 	&	62.49	&	Sy 1	&	1.437	$\pm$	0.073	&	2.416	$\pm$	0.137	&	13.78	&	12				&	22				&	1   	\\
MCG-01-33-063                 	&	13 00 19.1 	&	-08 05 15 	&	113.32	&	Sy 2 	&	0.302	$\pm$	0.027	&	1.160	$\pm$	0.062	&	10.32	&		350	9	23	&		350	15	33	&	1   	\\
SBS 1301+540                   	&	13 03 59.5 	&	+53 47 30 	&	129.23	&	Sy 1	&		$<$	0.151	&		$<$	0.085	&	32.61	&	12				&	22				&	1   	\\
NGC 4941                       	&	13 04 13.1 	&	-05 33 06 	&	18.22	&	Sy 2	&	0.965	$\pm$	0.058	&	3.955	$\pm$	0.216	&	20.45	&	12				&		16	41	71	&	1   	\\
NGC 4939                       	&	13 04 14.4 	&	-10 20 23 	&	44.16	&	Sy 2	&	1.938	$\pm$	0.123	&	7.000	$\pm$	0.379	&	24.85	&		0	49	72	&		0	49	72	&	1   	\\
ESO 323-077                    	&	13 06 26.1 	&	-40 24 53 	&	64.15	&	Sy 1.2	&	7.017	$\pm$	0.354	&	7.531	$\pm$	0.455	&	33.06	&	12				&	22				&	1   	\\
NGC 4992                       	&	13 09 05.6 	&	+11 38 03 	&	108.25	&	Sy 2 	&	0.180	$\pm$	0.010	&	0.548	$\pm$	0.047	&	53.47	&	5.5				&		0	21	30	&	1   	\\
II SZ 010                       	&	13 13 05.8 	&	-11 07 42 	&	148.60	&	Sy 1	&	0.085	$\pm$	0.006	&	0.125	$\pm$	0.013	&	15.20	&	12				&	22				&	1   	\\
NGC 5033                       	&	13 13 27.5 	&	+36 35 38 	&	19.64	&	Sy 1.9	&	19.133	$\pm$	0.959	&	50.152	$\pm$	2.534	&	5.52	&		170	31	68	&		170	54	83	&	1   	\\
UGC 08327NED02\tablenotemark{b}                 	&	13 15 17.3 	&	+44 24 26 	&	158.79	&	Sy 2	&	1.483	$\pm$	0.077	&	0.884	$\pm$	0.092	&	17.08	&	12				&	18				&	1   	\\
NGC 5106\tablenotemark{b}                       	&	13 20 59.6 	&	+08 58 42 	&	138.27	&	L	&	2.872	$\pm$	0.153	&	5.005	$\pm$	0.258	&	13.97	&	18				&	22				&	1   	\\
MCG-03-34-064                 	&	13 22 24.5 	&	-16 43 42 	&	70.76	&	Sy 1.8	&	5.099	$\pm$	0.257	&	3.444	$\pm$	0.198	&	30.53	&	12				&	22				&	1   	\\
Centaurus A	&	13 25 27.6 	&	-43 01 09 	&	3.66	&	Sy 2	&	400.576	$\pm$	20.031	&	732.884	$\pm$	36.647	&	1388.99	&		123	100	261	&		100	146	367	&	4	\\
ESO 509-G038                   	&	13 31 13.9 	&	-25 24 10 	&	111.91	&	Sy 1	&	0.441	$\pm$	0.023	&	0.732	$\pm$	0.079	&	14.24	&	12				&	22				&	1   	\\
ESO 383-18                     	&	13 33 26.1 	&	-34 00 53 	&	52.93	&	Sy 2	&	0.654	$\pm$	0.041	&	0.753	$\pm$	0.070	&	18.87	&	12				&	22				&	1   	\\
ESO 509-IG066NED01\tablenotemark{b}             	&	13 34 39.6 	&	-23 26 48 	&	148.78	&	Sy 2	&	0.883	$\pm$	0.048	&	1.656	$\pm$	0.154	&	18.18	&	10				&	26				&	1   	\\
NGC 5231                       	&	13 35 48.2 	&	+02 59 56 	&	93.46	&	Sy 2	&	0.702	$\pm$	0.039	&	1.771	$\pm$	0.125	&	15.62	&	12				&	22				&	1   	\\
MCG-06-30-015                 	&	13 35 53.7 	&	-34 17 44 	&	32.90	&	Sy 1.2	&	1.205	$\pm$	0.063	&	0.762	$\pm$	0.068	&	63.33	&	12				&	22				&	1   	\\
NGC 5252                       	&	13 38 16.0 	&	+04 32 33 	&	98.77	&	Sy 1.9	&	0.341	$\pm$	0.023	&	0.298	$\pm$	0.030	&	115.48	&	12				&	22				&	1   	\\
2MASX J13411287-1438407        	&	13 41 12.9 	&	-14 38 41 	&	182.26	&	Sy 1	&	0.132	$\pm$	0.009	&	0.198	$\pm$	0.022	&	24.07	&	12				&	16				&	1   	\\
NGC 5273                       	&	13 42 08.3 	&	+35 39 15 	&	15.96	&	Sy 1.9	&	0.721	$\pm$	0.040	&	0.705	$\pm$	0.063	&	13.91	&	12				&	22				&	1   	\\
CGCG 102-048                   	&	13 44 15.7 	&	+19 34 00 	&	116.73	&	Sy 1.9	&	0.052	$\pm$	0.008	&		$<$	0.147	&	19.51	&	5.5				&	10.5				&	1   	\\
NGC 5290                       	&	13 45 19.2 	&	+41 42 45 	&	34.98	&	Sy 2	&	2.071	$\pm$	0.240	&	7.339	$\pm$	1.290	&	19.17	&		272	19	76	&		272	19	76	&	1   	\\
4U 1344-60                     	&	13 47 36.0 	&	-60 37 04 	&	54.94	&	Sy 1.5	&	2.040	$\pm$	0.116	&		$<$	1.923	&	107.52	&	12				&	22\tablenotemark{a}				&	1   	\\
IC 4329A                       	&	13 49 19.3 	&	-30 18 34 	&	68.65	&	Sy 1.2	&	1.797	$\pm$	0.095	&	0.968	$\pm$	0.121	&	290.49	&	12				&	22				&	1   	\\
UM 614                         	&	13 49 52.8 	&	+02 04 45 	&	141.64	&	Sy 1	&		$<$	0.061	&		$<$	0.103	&	16.18	&	12				&	22				&	1   	\\
2MASX J13512953-1813468        	&	13 51 29.5 	&	-18 13 47 	&	52.09	&	Sy 1	&		$<$	0.076	&		$<$	0.079	&	17.88	&	12				&	22				&	1   	\\
Mrk 279\tablenotemark{b}                        	&	13 53 03.4 	&	+69 18 30 	&	131.67	&	Sy 1.5	&	0.852	$\pm$	0.044	&	0.822	$\pm$	0.073	&	39.20	&	12				&	22				&	1   	\\
ESO 578-G009                   	&	13 56 36.7 	&	-19 31 45 	&	151.96	&	Sy 1	&	0.318	$\pm$	0.030	&	0.999	$\pm$	0.075	&	16.90	&	18				&	22				&	1   	\\
2MASX J14080674-3023537        	&	14 08 06.8 	&	-30 23 54 	&	100.88	&	Sy 1.5	&		$<$	0.043	&		$<$	0.134	&	17.19	&	12				&	22				&	1   	\\
NGC 5506                       	&	14 13 14.9 	&	-03 12 27 	&	23.83	&	Sy 1.9	&	8.214	$\pm$	0.414	&	6.768	$\pm$	0.352	&	241.03	&		93	26	57	&		93	26	57	&	3	\\
NGC 5548                       	&	14 17 59.5 	&	+25 08 12 	&	73.51	&	Sy 1.5	&	1.173	$\pm$	0.064	&	1.494	$\pm$	0.098	&	78.64	&	12				&	22				&	1   	\\
ESO 511-G030                   	&	14 19 22.4 	&	-26 38 41 	&	96.21	&	Sy 1	&	0.405	$\pm$	0.010	&	2.021	$\pm$	0.132	&	39.88	&		160	34	54				&		160	34	54	&	1   	\\
NGC 5610                       	&	14 24 22.9 	&	+24 36 51 	&	72.26	&	Sy 2	&	5.419	$\pm$	0.275	&	6.219	$\pm$	0.315	&	19.21	&		270	23	34	&		270	23	34	&	1   	\\
NGC 5674                       	&	14 33 52.2 	&	+05 27 30 	&	107.35	&	Sy 1.9	&	1.819	$\pm$	0.098	&	3.940	$\pm$	0.202	&	15.97	&		0	34	39	&		0	34	39	&	1   	\\
NGC 5683                       	&	14 34 52.4 	&	+48 39 43 	&	157.31	&	Sy 1	&	0.097	$\pm$	0.010	&		$<$	0.169	&	12.27	&	5.5				&	10.5				&	1   	\\
Mrk 817                        	&	14 36 22.1 	&	+58 47 39 	&	136.11	&	Sy 1.5	&	2.159	$\pm$	0.110	&	1.706	$\pm$	0.154	&	25.91	&	12				&	22				&	1   	\\
Mrk 477\tablenotemark{b}                        	&	14 40 38.1 	&	+53 30 16 	&	164.03	&	Sy 1	&	1.307	$\pm$	0.070	&	0.735	$\pm$	0.047	&	14.40	&	12				&	10.5				&	1   	\\
NGC 5728                       	&	14 42 23.9 	&	-17 15 11 	&	30.57	&	Sy 2	&	9.758	$\pm$	0.489	&	13.717	$\pm$	0.687	&	88.63	&		30	38	75	&		30	38	75	&	3	\\
WKK 4374                       	&	14 51 33.1 	&	-55 40 38 	&	77.09	&	Sy 2	&	0.277	$\pm$	0.019	&		$<$	0.763	&	40.84	&	12				&	22\tablenotemark{a}				&	1   	\\
2MASX J14530794+2554327        	&	14 53 07.9 	&	+25 54 33 	&	203.52	&	Sy 1	&		$<$	0.131	&		$<$	0.129	&	24.01	&	12				&	22				&	1   	\\
WKK 4438                       	&	14 55 17.4 	&	-51 34 15 	&	68.42	&	Sy 1 	&	0.858	$\pm$	0.056	&	1.592	$\pm$	0.158	&	21.51	&	14				&	22				&	1   	\\
IC 4518A\tablenotemark{b}                       	&	14 57 41.2 	&	-43 07 56 	&	69.55	&	Sy 2	&	5.209	$\pm$	0.261	&	5.011	$\pm$	0.251	&	27.75	&	14				&	14				&	1   	\\
Mrk 841                        	&	15 04 01.2 	&	+10 26 16 	&	158.20	&	Sy 1	&	0.448	$\pm$	0.023	&	0.182	$\pm$	0.014	&	35.57	&	12				&	22				&	1   	\\
Mrk 1392                       	&	15 05 56.6 	&	+03 42 26 	&	156.92	&	Sy 1	&	0.364	$\pm$	0.026	&	0.936	$\pm$	0.063	&	19.01	&	12				&	22				&	1   	\\
2MASX J15064412+0351444        	&	15 06 44.1 	&	+03 51 44 	&	163.91	&	Sy 2	&	0.064	$\pm$	0.006	&		$<$	0.148	&	15.67	&	5.5				&	10.5				&	1   	\\
2MASX J15115979-2119015        	&	15 11 59.8 	&	-21 19 02 	&	194.95	&	Sy 1 	&	1.846	$\pm$	0.094	&	1.572	$\pm$	0.122	&	31.30	&	12				&	22				&	1   	\\
NGC 5899                       	&	15 15 03.2 	&	+42 02 59 	&	38.08	&	Sy 2	&	5.097	$\pm$	0.262	&	13.193	$\pm$	0.665	&	20.08	&		21	31	69	&		21	31	69	&	1   	\\
CGCG 319-007\tablenotemark{b}                   	&	15 19 33.7 	&	+65 35 59 	&	192.21	&	Sy 1.9	&	0.345	$\pm$	0.022	&	0.760	$\pm$	0.068	&	13.23	&	12				&	22				&	1   	\\
MCG-01-40-001                 	&	15 33 20.7 	&	-08 42 02 	&	97.60	&	Sy 2	&	2.056	$\pm$	0.108	&	2.425	$\pm$	0.125	&	32.72	&		78	21	40	&		78	21	40	&	1   	\\
Mrk 290                        	&	15 35 52.4 	&	+57 54 09 	&	127.80	&	Sy 1	&	0.178	$\pm$	0.010	&	0.116	$\pm$	0.008	&	23.31	&	12				&	22				&	1   	\\
2MASX J15462424+6929102\tablenotemark{b}        	&	15 46 24.3 	&	+69 29 10 	&	162.52	&	Sy 1.9	&	0.176	$\pm$	0.013	&	0.097	$\pm$	0.019	&	13.94	&	5.5				&	10.5				&	1   	\\
NGC 5995                       	&	15 48 25.0 	&	-13 45 28 	&	108.50	&	Sy 2	&	4.025	$\pm$	0.202	&	5.399	$\pm$	0.273	&	43.66	&	30				&	30				&	1   	\\
WKK 6092                       	&	16 11 51.4 	&	-60 37 55 	&	66.85	&	Sy 1	&	0.196	$\pm$	0.014	&		$<$	0.058	&	32.15	&	12				&	22				&	1   	\\
WKK 6471                       	&	16 18 36.4 	&	-59 27 17 	&	150.20	&	Sy 1	&	0.084	$\pm$	0.008	&	0.519	$\pm$	0.054	&	21.36	&	12				&	22				&	1   	\\
CGCG 367-009                   	&	16 19 19.3 	&	+81 02 48 	&	102.90	&	Sy 2	&	0.059	$\pm$	0.007	&	0.154	$\pm$	0.029	&	22.71	&	5.5				&	10.5				&	1   	\\
Mrk 885                        	&	16 29 48.3 	&	+67 22 42 	&	109.09	&	Sy 1	&	0.406	$\pm$	0.023	&	1.002	$\pm$	0.085	&	13.96	&	12				&	22				&	1   	\\
ESO 137-34                     	&	16 35 14.1 	&	-58 04 48 	&	32.95	&	Sy 2	&	2.793	$\pm$	0.150	&	6.220	$\pm$	0.386	&	28.19	&	53				&	53				&	1   	\\
2MASX J16481523-3035037        	&	16 48 15.3 	&	-30 35 04 	&	134.10	&	Sy 1	&	0.096	$\pm$	0.008	&		$<$	0.136	&	45.69	&	12				&	22				&	1   	\\
LEDA 214543                    	&	16 50 42.7 	&	+04 36 18 	&	138.79	&	Sy 2	&	0.118	$\pm$	0.008	&	0.326	$\pm$	0.022	&	24.49	&	12				&	22				&	1   	\\
UGC 10593                      	&	16 52 18.9 	&	+55 54 20 	&	125.92	&	Sy 2	&	0.510	$\pm$	0.035	&	0.917	$\pm$	0.064	&	15.15	&	12				&	22				&	1   	\\
NGC 6221                       	&	16 52 46.1 	&	-59 13 07 	&	12.34	&	Sy 2 	&	59.107	$\pm$	2.976	&	83.425	$\pm$	6.715	&	20.16	&		175	72	103	&		175	72	103	&	1   	\\
NGC 6240                       	&	16 52 58.9 	&	+02 24 03 	&	105.37	&	Sy 2	&	26.621	$\pm$	1.358	&	20.233	$\pm$	1.028	&	72.21	&	24				&	24				&	6   	\\
NGC 6300                       	&	17 16 59.5 	&	-62 49 14 	&	15.06	&	Sy 2	&	19.113	$\pm$	1.091	&	44.229	$\pm$	2.621	&	99.43	&		120	73	90	&		120	73	90	&	1   	\\
ARP 102B                       	&	17 19 14.5 	&	+48 58 49 	&	103.99	&	Sy 1 / L	&	0.103	$\pm$	0.009	&		$<$	0.133	&	18.20	&	5.5				&	10.5				&	1   	\\
AX J1737.4-2907                	&	17 37 28.4 	&	-29 08 03 	&	91.89	&	Sy 1	&	0.451	$\pm$	0.029	&		$<$	0.617	&	115.86	&	12				&	22\tablenotemark{a}				&	1   	\\
ESO 139-G012                   	&	17 37 39.1 	&	-59 56 27 	&	72.83	&	Sy 2	&	0.414	$\pm$	0.027	&	2.704	$\pm$	0.142	&	22.55	&	12				&		30	41	58	&	1   	\\
2E1739.1-1210                 	&	17 41 55.3 	&	-12 11 57 	&	160.78	&	Sy 1	&	0.561	$\pm$	0.028	&	0.781	$\pm$	0.114	&	44.50	&	12				&	22				&	1   	\\
CGCG 300-062                   	&	17 43 17.4 	&	+62 50 21 	&	142.96	&	Sy 2	&	0.102	$\pm$	0.006	&	0.286	$\pm$	0.021	&	10.08	&	12				&	22				&	1   	\\
2MASS J17485512-3254521        	&	17 48 55.1 	&	-32 54 52 	&	85.79	&	Sy 1	&		$<$	0.140	&		$<$	0.240	&	36.66	&	12				&	22				&	1   	\\
NGC 6552                       	&	18 00 07.3 	&	+66 36 54 	&	114.20	&	Sy 2	&	2.574	$\pm$	0.132	&	2.188	$\pm$	0.143	&	19.19	&	12				&	22				&	1   	\\
2MASXI J1802473-145454	&	18 02 47.3 	&	-14 54 55 	&	14.40	&	Sy 1	&	1.203	$\pm$	0.063	&	1.872	$\pm$	0.229	&	40.33	&	12				&	22				&	1   	\\
UGC 11185NED02\tablenotemark{b}                 	&	18 16 11.5 	&	+42 39 37 	&	179.63	&	Sy 2 	&	0.467	$\pm$	0.031	&	0.465	$\pm$	0.027	&	19.22	&	12				&	10.5				&	1   	\\
IC 4709                        	&	18 24 19.4 	&	-56 22 09 	&	72.34	&	Sy 2	&	0.592	$\pm$	0.040	&	0.943	$\pm$	0.051	&	42.55	&		0	13	28	&		0	20	50	&	1   	\\
Fairall 49                     	&	18 36 58.3 	&	-59 24 09 	&	85.88	&	Sy 2	&	3.318	$\pm$	0.169	&	2.555	$\pm$	0.147	&	14.16	&	12				&	22				&	1   	\\
ESO 103-035                    	&	18 38 20.3 	&	-65 25 39 	&	56.69	&	Sy 2	&	1.732	$\pm$	0.088	&	0.551	$\pm$	0.034	&	111.73	&	12				&	22				&	1   	\\
Fairall 51                     	&	18 44 54.0 	&	-62 21 53 	&	60.54	&	Sy 1	&	2.032	$\pm$	0.104	&	1.753	$\pm$	0.104	&	40.36	&	12				&	22				&	1   	\\
CGCG 341-006                   	&	18 45 26.2 	&	+72 11 02 	&	202.61	&	Sy 2	&	1.135	$\pm$	0.060	&	0.957	$\pm$	0.098	&	11.54	&	12				&	22				&	1   	\\
2MASX J18570768-7828212        	&	18 57 07.8 	&	-78 28 21 	&	183.20	&	Sy 1	&	0.349	$\pm$	0.019	&	0.356	$\pm$	0.031	&	24.22	&	12				&	22				&	1   	\\
CGCG 229-015                   	&	19 05 25.9 	&	+42 27 40 	&	120.31	&	Sy 1	&	0.148	$\pm$	0.009	&	0.270	$\pm$	0.022	&	11.14	&	12				&	22				&	1   	\\
ESO 141-G055                   	&	19 21 14.1 	&	-58 40 13 	&	161.27	&	Sy 1	&	0.613	$\pm$	0.041	&	1.310	$\pm$	0.088	&	54.42	&	12				&	22				&	1   	\\
2MASX J19373299-0613046        	&	19 37 33.0 	&	-06 13 05 	&	43.65	&	Sy 1.5	&	3.588	$\pm$	0.184	&	4.964	$\pm$	0.272	&	23.21	&	18				&	22				&	1   	\\
2MASX J19380437-5109497        	&	19 38 04.4 	&	-51 09 50 	&	174.30	&	Sy 1.2	&	0.078	$\pm$	0.010	&	0.211	$\pm$	0.024	&	15.05	&	5.5				&	10.5				&	1   	\\
NGC 6814                       	&	19 42 40.6 	&	-10 19 25 	&	22.80	&	Sy 1.5	&	7.224	$\pm$	0.370	&	16.265	$\pm$	0.824	&	77.27	&		96	55	58	&		96	55	58	&	1   	\\
2MASX J20005575-1810274        	&	20 00 55.7 	&	-18 10 27 	&	161.32	&	Sy 1 	&	0.886	$\pm$	0.046	&	0.696	$\pm$	0.062	&	20.90	&	12				&	22				&	1   	\\
ESO 399-20                     	&	20 06 57.7 	&	-34 32 58 	&	107.43	&	Sy 1 	&	0.622	$\pm$	0.040	&	1.441	$\pm$	0.074	&	17.93	&	26				&	26				&	1   	\\
NGC 6860                       	&	20 08 46.9 	&	-61 06 01 	&	63.59	&	Sy 1	&	1.431	$\pm$	0.076	&	3.059	$\pm$	0.160	&	53.18	&	22				&	26				&	3	\\
2MASX J20101740+4800214        	&	20 10 17.4 	&	+48 00 21 	&	110.41	&	Sy 2	&	0.198	$\pm$	0.012	&	0.352	$\pm$	0.033	&	13.26	&	12				&	22				&	1   	\\
2MASX J20183871+4041003        	&	20 18 38.7 	&	+40 41 00 	&	61.50	&	Sy 2	&		$<$	0.371	&		$<$	0.611	&	28.47	&	12\tablenotemark{a}				&	22\tablenotemark{a}				&	1   	\\
II Zw 083                       	&	20 26 55.9 	&	-02 16 39 	&	125.89	&	Sy 2	&	1.162	$\pm$	0.061	&	1.306	$\pm$	0.098	&	12.59	&	12				&	22				&	1   	\\
MCG+04-48-002\tablenotemark{b}                 	&	20 28 35.1 	&	+25 44 00 	&	59.34	&	Sy 2	&	9.698	$\pm$	0.486	&	10.445	$\pm$	0.524	&	77.79	&	60	22	33		&		60	22	33	&	1   	\\
Mrk 509                        	&	20 44 09.7 	&	-10 43 25 	&	149.18	&	Sy 1.2	&	1.501	$\pm$	0.078	&	1.230	$\pm$	0.088	&	96.16	&	12				&	22				&	1   	\\
IC 5063                        	&	20 52 02.3 	&	-57 04 08 	&	48.35	&	Sy 2	&	4.266	$\pm$	0.419	&	3.603	$\pm$	0.280	&	72.57	&		120	27	44	&		120	39	69	&	3	\\
ESO 464-G016                   	&	21 02 23.8 	&	-28 10 29 	&	157.89	&	AGN 	&	0.466	$\pm$	0.024	&	0.524	$\pm$	0.028	&	14.10	&	12				&	22				&	1   	\\
2MASX J21090996-0940147        	&	21 09 10.0 	&	-09 40 15 	&	114.30	&	Sy 1.2 / L	&	0.307	$\pm$	0.018	&	0.229	$\pm$	0.030	&	19.75	&	12				&	22				&	1   	\\
SWIFT J212745.6+565636         	&	21 27 44.9 	&	+56 56 40 	&	62.79	&	Sy 1	&	0.156	$\pm$	0.010	&		$<$	0.050	&	37.27	&	12				&	22				&	1   	\\
6dF J2132022-334254\tablenotemark{b}            	&	21 32 02.2 	&	-33 42 54 	&	129.56	&	Sy 1	&	0.079	$\pm$	0.007	&		$<$	0.082	&	42.69	&	5.5				&	10.5				&	1   	\\
2MASX J21355399+4728217        	&	21 35 54.0 	&	+47 28 22 	&	107.65	&	Sy 1	&	0.518	$\pm$	0.037	&	0.749	$\pm$	0.100	&	22.24	&	12				&	22				&	1   	\\
CGCG 493-002                   	&	21 38 33.4 	&	+32 05 06 	&	106.77	&	Sy 1.5	&	0.426	$\pm$	0.045	&	0.419	$\pm$	0.025	&	17.92	&	12				&	22				&	1   	\\
NGC 7172                       	&	22 02 01.9 	&	-31 52 11 	&	33.90	&	Sy 2	&	7.148	$\pm$	0.362	&	12.693	$\pm$	0.641	&	170.93	&		98	29	63	&		98	29	63	&	5	\\
NGC 7213                       	&	22 09 16.3 	&	-47 10 00 	&	22.00	&	Sy 1.5	&	3.237	$\pm$	0.165	&	7.961	$\pm$	0.401	&	42.10	&	50				&	50				&	1   	\\
MCG+02-57-002                 	&	22 23 45.0 	&	+11 50 09 	&	125.26	&	Sy 1.5	&	0.539	$\pm$	0.027	&	0.667	$\pm$	0.081	&	14.18	&	12				&	22				&	1   	\\
MCG+06-49-019                 	&	22 27 05.8 	&	+36 21 42 	&	91.58	&	Sy 2	&	0.142	$\pm$	0.013	&	0.517	$\pm$	0.039	&	19.77	&	5.5				&		80	17	28	&	1   	\\
ESO 533-G050                   	&	22 34 49.8 	&	-25 40 37 	&	114.00	&	Sy 2	&	0.103	$\pm$	0.015	&	0.398	$\pm$	0.029	&	13.55	&	5.5				&		20	18	29	&	1   	\\
MCG+01-57-016                 	&	22 40 17.0 	&	+08 03 14 	&	107.53	&	Sy 1.8	&	0.936	$\pm$	0.051	&	0.966	$\pm$	0.108	&	16.64	&	12				&	22				&	1   	\\
UGC 12237                      	&	22 54 19.7 	&	+11 46 57 	&	122.05	&	Sy 2	&	0.640	$\pm$	0.040	&	1.083	$\pm$	0.068	&	17.11	&	90	19	52		&		90	19	52	&	1   	\\
UGC 12282                      	&	22 58 55.5 	&	+40 55 53 	&	72.71	&	Sy 1.9	&	1.189	$\pm$	0.071	&	3.685	$\pm$	0.189	&	19.20	&		186	18	49	&		186	18	49	&	1   	\\
KAZ 320                        	&	22 59 32.9 	&	+24 55 06 	&	149.64	&	Sy 1	&	0.366	$\pm$	0.020	&	0.308	$\pm$	0.029	&	24.97	&	12				&	22				&	1   	\\
NGC 7465                       	&	23 02 01.0 	&	+15 57 53 	&	27.20	&	L	&	4.169	$\pm$	0.211	&	4.715	$\pm$	0.241	&	14.56	&	12				&	22				&	1   	\\
NGC 7469\tablenotemark{b}                       	&	23 03 15.6 	&	+08 52 26 	&	69.79	&	Sy 1.2	&	30.770	$\pm$	1.540	&	26.935	$\pm$	1.499	&	66.74	&	18				&	40				&	3	\\
Mrk 926                        	&	23 04 43.5 	&	-08 41 09 	&	205.15	&	Sy 1.5	&	0.606	$\pm$	0.033	&	0.713	$\pm$	0.067	&	113.89	&	12				&	22				&	1   	\\
NGC 7479                       	&	23 04 56.7 	&	+12 19 22 	&	33.85	&	Sy 2 / L	&	14.341	$\pm$	0.723	&	25.989	$\pm$	1.307	&	20.36	&		40	53	86	&		40	82	88	&	1   	\\
PG 2304+042\tablenotemark{b}                    	&	23 07 02.9 	&	+04 32 57 	&	183.20	&	Sy 1	&		$<$	0.025	&		$<$	0.028	&	15.00	&	12				&	22				&	1   	\\
NGC 7582                       	&	23 18 23.5 	&	-42 22 14 	&	20.62	&	Sy 2	&	67.409	$\pm$	3.393	&	77.261	$\pm$	3.967	&	80.65	&		156	61	123	&		156	61	123	&	5	\\
NGC 7603                       	&	23 18 56.6 	&	+00 14 38 	&	127.57	&	Sy 1.5	&	1.331	$\pm$	0.070	&	2.639	$\pm$	0.136	&	50.22	&	12				&	22				&	1   	\\
LCRS B232242.2-384320          	&	23 25 24.2 	&	-38 26 49 	&	155.87	&	Sy 1	&	0.644	$\pm$	0.037	&	1.339	$\pm$	0.077	&	14.97	&	12				&	22				&	1   	\\
2MASX J23272195+1524375        	&	23 27 22.0 	&	+15 24 37 	&	199.97	&	Sy 1	&	0.135	$\pm$	0.009	&	0.247	$\pm$	0.024	&	10.52	&	12				&	22				&	1   	\\
NGC 7679                       	&	23 28 46.7 	&	+03 30 41 	&	73.35	&	Sy 2	&	8.875	$\pm$	0.446	&	9.002	$\pm$	0.458	&	15.12	&	18				&	22				&	1   	\\
IGR J23308+7120                	&	23 30 37.7 	&	+71 22 46 	&	160.69	&	Sy 2	&	0.379	$\pm$	0.028	&	0.810	$\pm$	0.057	&	11.13	&	12				&	22				&	1   	\\
PKS 2331-240                   	&	23 33 55.2 	&	-23 43 41 	&	208.95	&	Sy 2	&	0.142	$\pm$	0.010	&	0.341	$\pm$	0.025	&	15.30	&	12				&	22				&	1   	\\
UGC 12741                      	&	23 41 55.5 	&	+30 34 54 	&	74.68	&	Sy 2	&	0.461	$\pm$	0.029	&	1.078	$\pm$	0.075	&	19.76	&	12				&	22				&	1   	\\

\enddata
\tablecomments{ Column 1: galaxy name. Column 2-3: coordinates. Column 4: luminosity distance in Mpc. To calculate the distance we   assumed a universe with a Hubble constant H$_o$ = 71 km s$^{–1}$ Mpc$^{–1}$, $\Omega_\Lambda$ = 0.73, and $\Omega_M$ = 0.27, with redshift values taken from NASA's ExtraGalactic Database (NED), except for sources with redshift values of z $<$ 0.01 where distances are taken from The Extragalactic Distance Database \citep[EDD, ][]{1988ngc..book.....T,2009AJ....138..323T}. Column 5: Galaxy type from \cite{2013ApJS..207...19B}, Seyfert galaxies (Sy), LINERs (L) and ESO 464-G016 \citep[AGN, unclassified Seyfert galaxy][]{2010A&A...518A..10V}. Column 6-7: PACS flux densities. Column 8: BAT fluxes in units of 10$^{-12}$ ergs s$^{-1}$ cm$^{-2}$  computed in the 14-195 keV band. Column 9-10. Aperture size in arcsec for PACS 70$\micron$ and 160$\micron$. For elliptical apertures there are three different values, the position angle (${\rm ^o}$) of the aperture's major axis measured from West to North and,  the minor and major diameter in arcsec. For point-like sources with flux densities fainter than $\sim$ 500 mJy we adopted smaller apertures as recommended by the NASA Herschel Science Center. Column 11: Herschel program ID, (1) OT1\_rmushotz\_1, (2) OT1\_lho\_1, (3) GT1\_msanchez\_2, (4) KPGT\_cwilso01\_1, (5) GT1\_lspinogl\_2, (6) KPGT\_esturm\_1, (7) OT2\_aalonsoh\_2. 
\label{table1}}
\tablenotetext{a}{High uncertainty due to cirrus contamination in the PACS field of view}
\tablenotetext{b}{There is at least one neighboring source inside the PACS field of view. In this case, we removed the neighboring source and selected  an aperture that contains only the emission from the BAT source. }

\end{deluxetable}

\section{Far-Infrared Properties of the BAT sample}
\label{bat}

\begin{figure}
\epsscale{.80}
\plotone{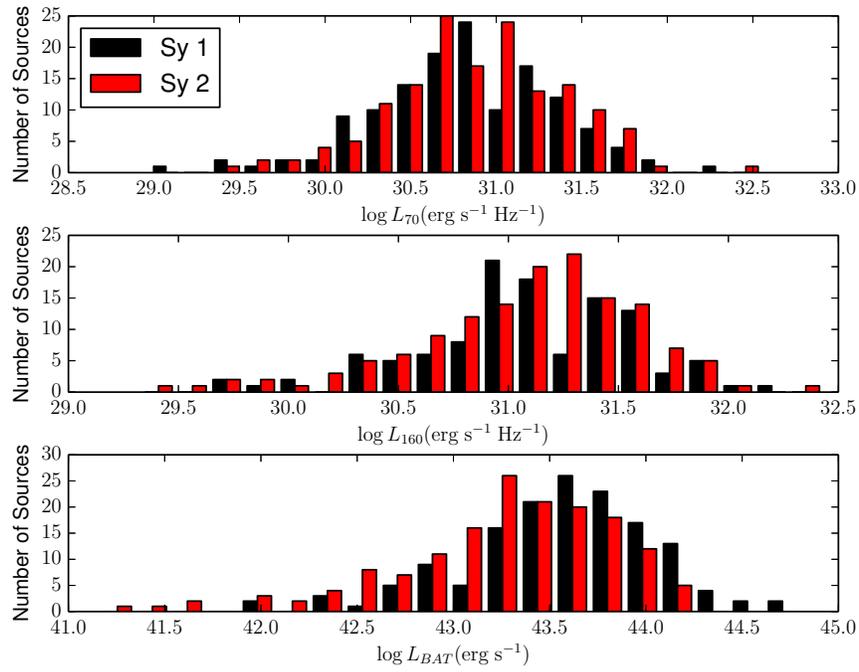}
\caption{The distributions of the monochromatic luminosities  for the emission at 70 and 160~$\micron$ and the integrated 14-195 keV luminosities for the sample of BAT AGN presented in this work.  Upper limits are not included in the distribution.   \label{fig1}}
\end{figure}

Figure~\ref{fig1} presents the histograms comparing the   70~$\micron$, 160~$\micron$ and BAT 14-195 keV luminosities of the Seyfert 1 and Seyfert 2 galaxies. We performed a  Kolmogorov-Smirnov (K-S) test  on the detected sources   to determine whether the BAT  Seyfert 1 and 2 galaxy  populations are drawn from the same parent distribution. A K-S probability value of less than 5$\%$ is the probability that two samples drawn from the same parent population would differ this much 5$\%$ of the time, i.e., that they are statistically different. A strong level of significance is obtained for values smaller than 1$\%$ \citep[e.g., ][]{1992nrfa.book.....P,2003drea.book.....B}. The number of sources, mean values, standard deviations  and the K-S probability of the null hypothesis for the sample are presented in Table~\ref{table2}. The K-S test for the luminosity  distribution at 70~$\micron$  returns a 70.0$\%$ probability of the null hypothesis (i.e., the  Seyfert 1 and Seyfert 2 galaxy populations are not distinguished by the present data). Similarly, the K-S null probability for the luminosity  distribution at 160~$\micron$  is  52.0$\%$. A similar situation for the Seyfert galaxies in the BAT sample  has also been found in the luminosities  of MIR  narrow emission lines \citep{2010ApJ...716.1151W} and of optical, reddening-corrected  emission lines \citep{2010ApJ...710..503W}. 

Caution must be taken when applying statistical tests to data sets that contain non-detections (upper limits), or ``censored" data points. In order to deal with these problems, we used the Astronomy SURVival analysis software package ASURV Rev 1.2  \citep{1986ApJ...306..490I,1992ASPC...25..245L}.  We performed statistical two-sample tests  and found that  the luminosity  distribution at 70~$\micron$  returns a 23.8$\%$ probability  when using Gehan's Generalized Wilcoxon test, in other words, two samples drawn from the same parent population would differ this much 23.8$\%$ of the time. This result is in agreement, within the statistical significance,  with the K-S test on the detected sources.  On the other hand,  the luminosity  distribution at 160~$\micron$  returns a Gehan's Generalized Wilcoxon test probability of 5.3$\%$. A similar two-sample test for censored data, the Peto-Prentice generalized Wilcoxon test, returns a probability of only 4.2 $\%$ that the Seyfert 1's and 2's  populations at 160~$\micron$ are drawn from the same parent population. Thus, the luminosity  distributions of the  Seyfert 1 and 2 galaxy   populations  at 160~$\micron$  differ at a weak but possibly statistically significant  level when non-detections are included in the analysis. From this we find that Seyfert 2 galaxies have a very slightly higher mean luminosity at 160~$\micron$, $\log$L$_{160\micron} = 30.98 \pm 0.05$ (erg s$^{-1}$ Hz$^{-1}$), than Seyfert 1 galaxies, $\log$L$_{160\micron} = 30.72 \pm 0.08$ (erg s$^{-1}$ Hz$^{-1}$). By comparison, the six sources uniquely identified as LINERs in our sample have  mean luminosities at 70 and 160~$\micron$ of  $\log$L$_{70\micron} = 30.70 \pm 0.34$ (erg s$^{-1}$ Hz$^{-1}$) and $\log$L$_{160\micron} = 30.92  \pm 0.29$ (erg s$^{-1}$ Hz$^{-1}$), respectively, thus different to Seyfert 1 and Seyfert 2 galaxies. The number of sources with detected and ``censored" data points and the Gehan's Generalized Wilcoxon test probability of the null hypothesis for the sample are presented in Table~\ref{table2}. In addition, Table~\ref{table2} also shows the  mean value for the entire sample of Seyfert 1 and 2   galaxies (including detections and non-detections) based  on the    Kaplan-Meier estimator  with  randomly censored data \citep{doi:10.1080/01621459.1958.10501452}.


 The similar FIR luminosities  in these two types of  Seyfert galaxies  implies  that radiation at these wavelengths is roughly isotropic, e.g.,  independent of orientation. This result suggests  two plausible  scenarios: (i) star formation is isotropic, and hence the FIR luminosities of Seyfert 1 and 2 galaxies would be indistinguishable and/or (ii) the  AGN torus is isotropic at FIR wavelengths.  In the former scenario,  we find that a large number of BAT sources are point-like in the PACS 70~$\micron$ images, implying a compact FIR nucleus of less than 6\arcsec (which typically encompasses regions of less than a couple of kpc). Thus, if star formation is the culprit of most of the FIR emission, then  it has to happen in a very compact, nuclear region suggesting a connection between the AGN and the nuclear cold molecular gas \citep[see][]{2014ApJ...781L..34M}. The latter scenario is in agreement with some theoretical calculations of the dusty, obscuring material surrounding the accretion disk that predict a similar shape for the FIR part of the SED for both types of galaxy populations \citep[e.g.,][]{2003PASP..115..928K,2012MNRAS.420.2756S}. Note that some torus  models predict an angle dependence on the FIR luminosity, with  Seyfert 1 galaxies having higher luminosities than Seyfert 2 galaxies \citep[e.g., ][]{2006MNRAS.366..767F}.  Thus, in order to match the observed FIR part of the SED,   predictions from these models require an extra contribution from a circumnuclear starburst in edge-on systems. In general, much of the light would have its origin in star formation.

On the other hand, the 14-195 keV X-ray luminosities (Figure~\ref{fig1}), bottom panel) are statistically different between the Seyfert types with a K-S test probability of 8.0 $\times 10^{-3} \%$, e.g, two samples drawn from the same parent population  would differ this much 8.0 $\times 10^{-3} \%$ of the time. We find that  Seyfert 2 galaxies have, on average,  smaller luminosities, $\log$L$_{BAT} = 43.26 \pm 0.04$ (erg s$^{-1}$), than Seyfert 1 galaxies with  $\log$L$_{BAT} = 43.57 \pm 0.05$ (erg s$^{-1}$), in agreement with previous results \citep{2008ApJ...682...94M,2009ApJ...690.1322W,2010ApJ...716.1151W,2011ApJ...728...58B,2012ApJ...754...45I}. This result suggests two possible scenarios: (1)  Compton down scattering, even in the high energy  14-195 keV band,   for some of the Seyfert 2 galaxies may be important in reducing the observed flux \citep[e.g.,][]{2009ApJ...692..608I,2009MNRAS.397.1549M}, or (2) the statistical differences between absorbed (Seyfert 2s) and unabsorbed (Seyfert 1s) AGN  are   more fundamental  with  absorbed AGN being  intrinsically less luminous than  unabsorbed AGN in agreement with previous studies \citep[e.g.,][]{2003ApJ...584L..57C,2003ApJ...596L..23S,2003ApJ...598..886U,2005AJ....129..578B,2005ApJ...635..864L,2005A&A...441..417H,2011ApJ...728...58B}. Regarding the former scenario, one must consider that there is no observational evidence of a dominant  population of Compton thick (CT) AGN in the BAT survey \citep[e.g.,][]{2011ApJ...728...58B}. Moreover,  the fraction of CT objects in deep hard X-ray surveys has been estimated with some precision to be only $\sim$ 17$\%$ \citep[e.g.,][]{2013arXiv1302.2451B}.

The latter scenario, an intrinsic difference between the luminosities of Seyfert types,  is supported by the difference in the  luminosity break found in the luminosity function between  the two classes of objects for the BAT sample, with absorbed AGN having (on average)  lower luminosities than unabsorbed AGN \citep[e.g.,][]{2011ApJ...728...58B}.  Overall, this scenario provides a test for the basic predictions from the unified model of AGN in which  the intrinsic AGN luminosity should be independent of obscuration; that is, one must consider   a model where the physical properties of the dusty, molecular torus change as a function of the intrinsic properties of the AGN central engine, e.g., accretion rate, power, etc \citep[e.g.,][]{1991MNRAS.252..586L}. For the purpose of this paper, we shall consider both scenarios as plausible and discuss their implications in the context of the FIR emission of AGN.

\begin{deluxetable}{lccccccc}
\tablecolumns{8}
\tabletypesize{\scriptsize}
\rotate
\tablewidth{0pt}
\tablecaption{Statistical Analysis for the Sample}
\tablehead{
\colhead{Observable}           & \colhead{ Measurements}      & \colhead{Mean}           & \colhead{Standard}      & 
\colhead{ Measurements}      & \colhead{Mean}           & \colhead{Standard}      & \colhead{$p$}\\
\colhead{ }           & \colhead{Available}      &         & \colhead{Deviation}      & 
\colhead{ Available}      &   & \colhead{Deviation}      & \\
\tableline

\colhead{(1)} & \colhead{(2)} & \colhead{(3)}&
\colhead{(4)} & \colhead{(5)} & \colhead{(6)} &
\colhead{(7)} & \colhead{(8)}\\
 & \multicolumn{3}{c}{(Seyfert 1)}&  \multicolumn{3}{c}{(Seyfert 2)} \\

}
\startdata

$\log$ L$_{70\micron}$	&	137	&	30.83	&	0.05	&	151	&	30.88	&	0.04	&	7.0 $\times 10^{-1}$	\\
$\log$ L$_{160\micron}$	&	113	&	31.08	&	0.04	&	141	&	31.07	&	0.05	&	5.2 $\times 10^{-1}$\\
$\log$ L$_{{\rm BAT}}$	&	149	&	43.56	&	0.04	&	157	&	43.26	&	0.05	&	8.0 $\times 10^{-5}$\\
L$_{70\micron}$/L$_{160\micron}$ & 113 & 0.81 & 0.43 & 139 & 0.77 & 0.47 & 5.4 $\times 10^{-1}$ \\
\cutinhead{Statistical Analysis (censored data)}
$\log$ L$_{70\micron}$\tablenotemark{a}	&	149 (12)	&	30.73	&	0.05	&	157 (6)	&	30.84	&	0.04	&	2.4$\times 10^{-1}$	\\
$\log$ L$_{160\micron}$\tablenotemark{a}&       149 (36)&	30.72	&	0.08 	&  	157 (16) 	&  30.98	 	&	0.05 	& 5.0 $\times 10^{-2}$ 	\\

\enddata
\tablecomments{Column 1: the observed data used for the correlation analysis. Column 2: number of Seyfert 1 galaxies used for the correlation analysis. Column 3: mean value for Seyfert 1 galaxies. Column 4: standard deviation for Seyfert 1 galaxies. Columns (5), (6) and (7) are the same as columns (2), (3) and (4) but for Seyfert 2 galaxies. Column (8):  the K-S test null probability for the detected data points and  Gehan's Generalised Wilcoxon test probability when upper limits are considered.  In order to calculate the mean value and standard deviations  with ``censored" data points (number of sources with upper limits in parentheses),  we used the Kaplan-Meier estimator  with  randomly censored data \citep{doi:10.1080/01621459.1958.10501452}. }
\tablenotetext{a}{ The mean values and standard deviations are given by the Kaplan-Meier estimator. The two-sample test probability is given by  Gehan's Generalized Wilcoxon test. \label{table2}}

\end{deluxetable}

\section{Far-Infrared and Hard X-ray Correlations}
\label{corre}

\begin{figure}
\epsscale{.80}
\plotone{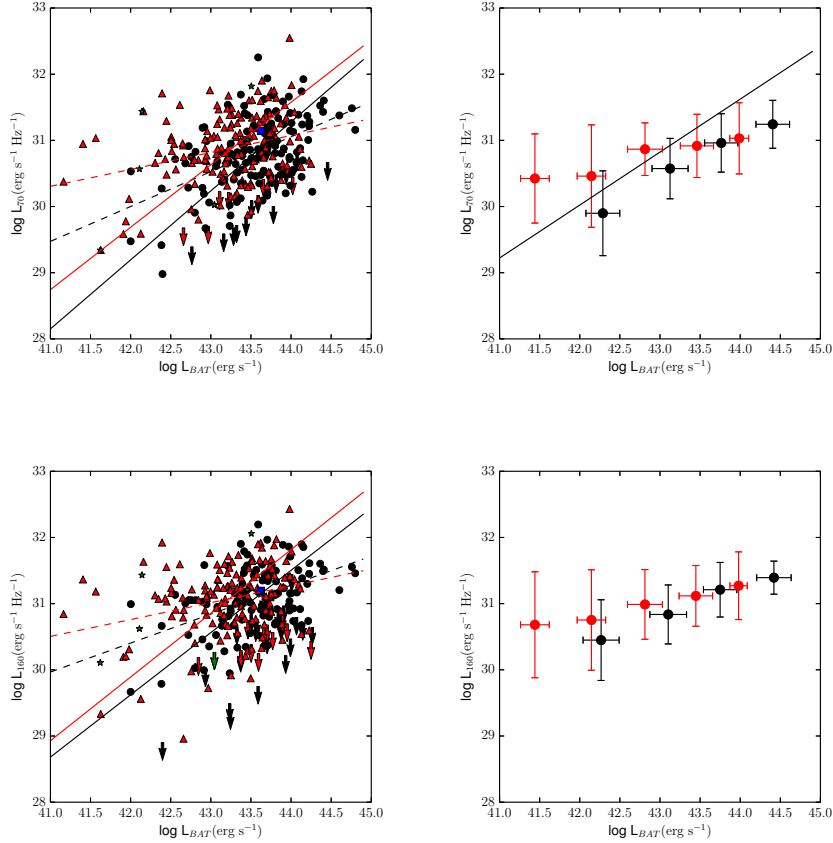}
\caption{Left panel: correlation between the monochromatic luminosities at 70 and 160~$\micron$  and the 14-195 keV luminosity (L$_{BAT}$) in Seyfert 1 (black circles), Seyfert 2 (red triangles), LINERs (green stars) and AGN (blue square). AGN refers to unidentified or previously unknown AGN. The dashed and solid lines represent the linear regression using  the OLS and bisector methods, respectively, separated into  Seyfert 1 (black lines) and Seyfert 2 (red lines) galaxies. Right panel: binned correlation  between  the monochromatic luminosity at 70 and 160~$\micron$  and the 14-195 keV luminosity  in Seyfert 1 (black) and Seyfert 2 (red) galaxies. We choose 5 equally spaced bins within the range of the FIR monochromatic and BAT luminosities. The error bars are the standard deviations for each quantity. The solid  line in the upper right panel is the AGN dominated line from \cite{2009MNRAS.399.1907N} (see text for details).  \label{fig2}}
\end{figure}

As we mentioned before, FIR emission is widely used as a probe of star formation. However, it is clear that if the AGN has some contribution at  FIR  wavelengths, then  one must correct  these SFR indicators accordingly. Figure~\ref{fig2} shows 14-195 keV luminosities   versus the  70~$\micron$ or 160~$\micron$  luminosity.  Due to redshift effects, luminosity--luminosity plots will almost always show some correlation. So,   we used a non-parametric test for partial correlation with censored data \citep{1996MNRAS.278..919A} in order to exclude the redshift effect. From this we find that the 14-195 keV luminosities are statistically correlated with the FIR luminosities  in the Seyfert 1 galaxies with a partial Kendall $\tau_p$=0.201 and a probability of $P_{\tau} = 5.3\times 10^{-6}$ at 70~$\micron$,  and $\tau_p$=0.116 and $P_{\tau} = 6.7\times 10^{-3}$ at 160~$\micron$. We find no statistically significant correlation between the 14-195 keV and FIR luminosities in the Seyfert 2 galaxies; in other words, the 14-195 keV and FIR luminosity distributions are independent (see Table~\ref{tablek} for details). In order to test whether these correlations are dependent on the BAT luminosity, we used the partial Kendall test in  two different groups: sources with $\log L_{BAT} > 43.0 $ ($\log L_{BAT} > 42.5 $) and $\log L_{BAT} > 44.0 $ ($\log L_{BAT} > 43.5 $) for Seyfert 1 (Seyfert 2) galaxies. For Seyfert 1 galaxies at 70~$\micron$, we find $\tau_p$=0.189 and  $P_{\tau} = 1.6\times 10^{-4} $ ($\log L_{BAT} > 43.0 $), and $\tau_p$=0.258 and $P_{\tau} = 3.0\times 10^{-2}$ ($\log L_{BAT} > 44.0 $). Similarly,  at 160~$\micron$ we find $\tau_p$=0.117 and $P_{\tau} = 1.8\times 10^{-2}$ ($\log L_{BAT} > 43.0 $), and  $\tau_p$=0.226 and $P_{\tau} = 2.3\times 10^{-2}$ ($\log L_{BAT} > 44.0 $). Again, we find no statistically significant correlation between the 14-195 keV and FIR luminosities in the Seyfert 2 galaxies   within the luminosity groups (see Table~\ref{tablek} for details). Note that in the Seyfert 1 galaxies the FIR-X-ray correlations get stronger for the most X-ray  luminous objects suggesting that the AGN contribution overwhelms the SF contribution at high  luminosities.  

\begin{deluxetable}{lcccccc}
\tablecolumns{7}
\tabletypesize{\scriptsize}
\tablewidth{0pt}
\tablecaption{Partial Correlation Analysis for the Sample}
\tablehead{
\colhead{$\log X$ vs $\log Y$}            & \colhead{$\tau_p$}      & \colhead{$\sigma$}           & \colhead{$P_{\tau}$}      & 
\colhead{$\tau_p$}      & \colhead{$\sigma$}           & \colhead{$P_{\tau}$}\\

\tableline

\colhead{(1)} & \colhead{(2)} & \colhead{(3)}&
\colhead{(4)} & \colhead{(5)} & \colhead{(6)} &
\colhead{(7)} \\
 & \multicolumn{3}{c}{(Seyfert 1)}&  \multicolumn{3}{c}{(Seyfert 2)} \\

}
\startdata

L$_{{\rm BAT}}$ - L$_{70}$ & 0.201 &	0.044    & 5.3$\times 10^{-6}$ &	 0.076	& 0.041&  6.4$\times 10^{-2}$	\\
L$_{{\rm BAT}}$ - L$_{160}$ &0.116 &	0.043	 & 6.7$\times 10^{-3}$  & -0.005	 	& 0.042& 9.1$\times 10^{-1}$	\\
 &\multicolumn{3}{c}{$\log L_{BAT} > 43.0 $}&  \multicolumn{3}{c}{$\log L_{BAT} > 42.5 $} \\
L$_{{\rm BAT}}$ - L$_{70}$ & 0.189 & 0.050 & 1.6$\times 10^{-4}$ & 	0.063 & 0.045 & 1.6$\times 10^{-1}$\\
L$_{{\rm BAT}}$ - L$_{160}$ & 0.117 & 0.049 & 1.8$\times 10^{-2}$ &  -0.007 & 0.046 & 8.7$\times 10^{-1}$ \\

&\multicolumn{3}{c}{$\log L_{BAT} > 44.0 $}&  \multicolumn{3}{c}{$\log L_{BAT} > 43.5 $} \\
L$_{{\rm BAT}}$ - L$_{70}$ & 0.258 & 0.119 & 3.0$\times 10^{-2}$ & 	0.004 & 0.090 & 9.6$\times 10^{-1}$ \\
L$_{{\rm BAT}}$ - L$_{160}$ & 0.226 &  0.100 & 2.3$\times 10^{-2}$ & 0.009 & 0.096 & 9.2$\times 10^{-1}$ \\

\sidehead{W/O CT} 
L$_{{\rm BAT}}$ - L$_{70}$ &   &	    &   & 0.065 & 0.043& 1.2$\times 10^{-1}$	\\
L$_{{\rm BAT}}$ - L$_{160}$ &   &	    &   & -0.014 & 0.044& 7.5$\times 10^{-1}$	\\
\sidehead{W/O CT and Candidates for CT} 
L$_{{\rm BAT}}$ - L$_{70}$ &   &	    &   & 0.084 & 0.048& 8.0$\times 10^{-2}$	\\
L$_{{\rm BAT}}$ - L$_{160}$ &   &	    &   & -0.019 & 0.049& 9.6$\times 10^{-1}$	\\

\enddata
\label{tablek}
\tablecomments{Column 1:  log X and log Y represent the independent and dependent variables, respectively. Column 2: $\tau_p$ is the Kendall's coefficient for partial correlation with censored data for Seyfert 1 galaxies. Column 3: $\sigma$ is the calculated variance for Seyfert 1 galaxies. Column 4: $P_{\tau}$ is the  associated null probability for Kendall $\tau_p$ for Seyfert 1 galaxies. Columns (5), (6) and (7) are the same as columns (2), (3) and (4) but for Seyfert 2 galaxies. We also show the correlation analysis for the sample of Seyfert 2 galaxies when Compton Thick (CT, Table~\ref{CT}) sources are excluded (W/O CT) and, when CT and candidates for CT  are excluded (W/O CT and Candidates for CT, see Table~\ref{CT} and \ref{CC}).  }

\end{deluxetable}

Overall, for the Seyfert 1 galaxies, we find a better correlation at 70~$\micron$  than at 160~$\micron$. This result is in agreement  with previous works  where the correlations between  14-195 keV luminosities   and  different monochromatic infrared luminosities get weaker at  longer wavelengths where the contribution from star formation might be  greater. For example,  the tightest and most significant correlations are  found   between the 14-195 keV luminosities and the 9,12 and 18 $\micron$ emission \citep[e.g., ][]{2009A&A...502..457G,2012ApJ...753..104M,2012ApJ...754...45I}, and the correlations get less significant  at longer   FIR wavelengths \citep[e.g., ][]{2008ApJ...689...95M,2008ApJ...685..160N,2012ApJ...753..104M,2012ApJ...754...45I}. However, this is the first time that a weak but possibly statistically significant correlation between the intrinsic power of the AGN and the FIR emission at 160~$\micron$ has been found. These results suggest two possible scenarios: (1) if we assume  that the FIR luminosity is a good tracer of star formation \citep[e.g.,][]{2010ApJ...714.1256C}, then there is a connection between star formation and the AGN at sub-kiloparsec scales \citep[e.g.,][]{2014ApJ...781L..34M}, or (2) dust heated by the  AGN has a statistically significant  contribution to the FIR emission at 70 and 160\micron. In the latter scenario,   SFR indicators that rely on FIR emission, either through the individual infrared bands \citep[][]{2010ApJ...714.1256C} or the total FIR emission \citep[8--1000 $\micron$, ][]{1998ARA&A..36..189K}, need to consider the AGN contribution in their predictions.  Note that in order to  increase the FIR emission predicted from the outer (and colder) regions of many torus models (AGN contribution), one could use any  combination of free parameters, such as an  increase of the torus radius with a constant optical depth, a flatter radial density profile  and/or  an  edge-on orientation \citep[e.g., ][]{2008ApJ...685..160N,2012MNRAS.420.2756S}. Within these parameters (uncertainties), smooth, continuous torus models seem to be able to predict higher FIR fluxes than clumpy torus calculations (with a broader range of SED shapes); however,  clumpy models provide a better match to the  MIR portion of the SED \citep[e.g.,][]{2011MNRAS.414.1082M}.

In addition,  Figure~\ref{fig2}  shows different linear regression methods applied to the sample (see Table~\ref{table3} for values). In order to test the effect of the upper limits in our sample, we estimated the linear regression coefficients using the EM (expectation-maximization) algorithm with censored data \citep{1986ApJ...306..490I} and an ordinary least-square regression of the dependent variable, Y, against the independent variable X, (OLS)  without the inclusion of censored data (only detections). Both methods show a good agreement, within their uncertainties,  suggesting that the non-detections  in our sample do    not  significantly change the results. However, it is not clear that the FIR luminosity is a direct  consequence of the BAT luminosity; thus,  a bisector method may be  more appropriate to investigate the underlying  functional   relationship between the hard X-ray and FIR luminosities \citep{1990ApJ...364..104I}. From this we find a nearly  linear relationship between the AGN and PACS monochromatic luminosities,  with Seyfert 1 and Seyfert 2 galaxies having similar slopes within the given uncertainties. Overall, these values are in agreement  with the  slopes  found by \cite{2012ApJ...753..104M} between the 14-195 keV and 90~$\micron$ luminosities.

\begin{deluxetable}{lcccccccccc}
\tabletypesize{\scriptsize}
\rotate
\tablewidth{0pt}
\tablecaption{Linear regressions for the Far-infrared and the 14--195 keV Luminosities}
\tablehead{
\colhead{$\log X$ vs $\log Y$}           & \multicolumn{2}{c}{$\log Y=m\log X+b$\tablenotemark{1}}& \multicolumn{2}{c}{$\log Y=m\log X+b$\tablenotemark{1}}& \multicolumn{2}{c}{$\log Y=m\log X+b$\tablenotemark{1}}\\
\tableline
                                         & m & b                     &  m & b                    &  m & b                     \\
\tableline
                \colhead{(1)}           & \multicolumn{2}{c}{(2)}     & \multicolumn{2}{c}{(3)}     & \multicolumn{2}{c}{(4)}\\     
}
\startdata
\sidehead{Seyfert 1 Galaxies:} 
L$_{{\rm BAT}}$ - L$_{70}$& 0.54  $\pm$   0.09  & 7.45   $\pm$   3.79 &  0.52   $\pm$     0.08  &  8.14   $\pm$    3.70  &  1.04    $\pm$    0.07   &   -14.60   $\pm$   2.87 \\
L$_{{\rm BAT}}$ - L$_{160}$& 0.45    $\pm$     0.11 & 11.06   $\pm$   4.74  &   0.44    $\pm$     0.08   &    11.95    $\pm$    3.51  &    0.94   $\pm$     0.06  &   -9.88    $\pm$      2.74 \\
\sidehead{Seyfert 2 Galaxies:} 
L$_{{\rm BAT}}$ - L$_{70}$& 0.26    $\pm$    0.07  &   19.83   $\pm$    3.22  & 0.25    $\pm$    0.08   &    20.13     $\pm$     3.64      &    0.94    $\pm$    0.04    &    -9.96    $\pm$     1.84 \\
L$_{{\rm BAT}}$ - L$_{160}$& 0.13    $\pm$    0.08 & 25.30     $\pm$     3.33  & 0.25     $\pm$      0.09   &    20.32    $\pm$     4.06     &   0.96     $\pm$    0.05    &   -10.52     $\pm$    2.09  \\

\enddata
\tablenotetext{1}{ m and b represent the regression coefficient (slope) and regression constant (intercept), respectively.}
\tablecomments{Column 1:  log X and log Y represent the independent and dependent variables, respectively. Column 2: linear regression coefficients using the EM  algorithm with censored data (upper limits). Column 3: ordinary least-square regression of the dependent variable, Y, against the independent variable X, OLS(Y|X) without the inclusion of censored data. Column 4: OLS bisector method without censored data which treat the variables symmetrically. See \cite{1990ApJ...364..104I} for a review of all the methods. \label{table3} }

\end{deluxetable}

Included as well in Figure~\ref{fig2} is the relation between $L_{60}$ and $L_{bol}$ for AGN dominated sources from \cite{2009MNRAS.399.1907N} (black line in upper right panel). In order to compare this relation with our sample, we assumed a  mean ratio of $S_{70}/S_{60}= 1.09$, as derived from the PACS-IRAS comparison presented in Figure~\ref{PACS_IRAS}, and a constant ratio of 10.5  to transfer the 14-195 keV luminosity to bolometric luminosity \citep{2012ApJ...745..107W}. Overall, the \cite{2009MNRAS.399.1907N}  relationship for AGN dominated sources is in fair agreement with the mean values of our sample, and it extends linearly over  our range of  X-ray luminosities. Note that systems above this straight line  may be dominated by star formation in the host galaxy resulting in the  flatter slope observed in Seyfert 2 galaxies  at low BAT luminosities, $\log L_{BAT} < \sim42.5$ erg s$^{-1}$ (or $\log L_{bol} < \sim43.7$ erg s$^{-1}$). This luminosity is smaller, by a factor of $\sim$ 3, than the turnover value found by \cite{2012A&A...545A..45R} for local AGN, $\log L_{bol} = 44.23\pm 0.13$ (erg s$^{-1}$), however, Seyfert 1 galaxies show no turnover over our range of X-ray luminosities. \cite{2009MNRAS.399.1907N} suggested a time evolution scenario to explain  the $L_{60}$ versus $L_{bol}$   linear correlation, where the galaxy transitions  from a pure  starburst to  a powerful composite starburst-AGN and, finally, to a weak composite starburst-AGN phase. In this scenario, after a  long star formation period, some of the cold gas is funneled to  the center of the galaxy and feeds the black hole, resulting in  a short period of intense AGN and stellar  activity  with  high star formation and accretion rate. This stage is characterized by sources rising and moving horizontally  above the straight line in the right upper panel in Figure~\ref{fig2}. As the cold  gas supply diminishes, both the star formation and the central AGN fade in parallel and fall below the straight  line \cite[see Figure~14 in ][]{2009MNRAS.399.1907N}. Interestingly, this scenario may  suggest an analogous  evolutionary connection between the Seyfert 1 and Seyfert 2 branches in our  L$_{70}$ versus L$_{BAT}$ comparison. Here Seyfert 2 galaxies, dominated by star formation,  are above the straight line. As they  evolve in time,  they reach  their maximum AGN and star formation activity until the  Seyfert 1 and Seyfert 2 galaxy branches   grow  together at high star formation, L$_{70}$, and AGN luminosity, L$_{BAT}$. As the supply of  cold gas is reduced, the AGN and  star formation luminosity fade together below the straight line, evolving into the  observed Seyfert 1 galaxy branch.   Note that  the slope of the \cite{2009MNRAS.399.1907N}  relationship ($L_{60} \propto L_{bol}^{0.8}$)    is in good agreement with our  values derived from the bisector method whereas the OLS slopes are much flatter.

\subsection{Seyfert 2 Galaxies}

In the previous section, we proposed two scenarios to explain the statistical differences in the 14-195 keV luminosity distribution between Seyfert 1 and Seyfert 2 galaxies: the effect of Compton scattering in the Seyfert 2 population,  or  that Seyfert 2 galaxies  are  intrinsically less luminous than Seyfert 1 galaxies.  In the following discussion, we will  investigate the effect of these scenarios in the FIR and hard X-ray relationship in order to find  the culprit for the lack of correlation between the 14-195 keV and FIR luminosities in Seyfert 2 galaxies. In the former scenario, if Compton scattering in the 14-195~keV band is important, then the more heavily obscured sources are responsible for breaking the intrinsic correlation between the FIR and the hard X-ray by reducing the X-ray flux and, perhaps, increasing the FIR emission. This scenario is in agreement with the  lack of a correlation found between the 14-195~keV luminosities  and the {\it AKARI}  infrared  luminosities  at 90~$\micron$ in CT AGN for the the 22-month BAT survey \citep{2012ApJ...753..104M}. Moreover, in this sample, Seyfert 2 galaxies  are correlated if CT AGN are excluded. In order to test the effect of high X-ray column densities on the FIR and X-ray correlations,   we have compiled published values for the X-ray column densities of Seyfert 2 galaxies in our sample \citep[see][and references therein]{2008ApJ...681..113T,2009ApJ...690.1322W,2011ApJ...728...58B,2013ApJ...763..111V}. From this we found that 106 out of 157 Seyfert 2 galaxies have  published X-ray column densities with 18 sources having high X-ray column densities, ${\rm N_H > 10^{24}}$~${\rm cm^{-2}}$ (see Table~\ref{CT}). We use the same partial correlation analysis as before and find that, even when excluding the CT sources presented in Table~\ref{CT},  there is no statistically significant correlation between the 14-195 keV and FIR luminosities (70 and 160~$\micron$) in Seyfert 2 galaxies when the influence of  distance is excluded (see Table\ref{tablek}).

\begin{figure}
\epsscale{.80}
\plotone{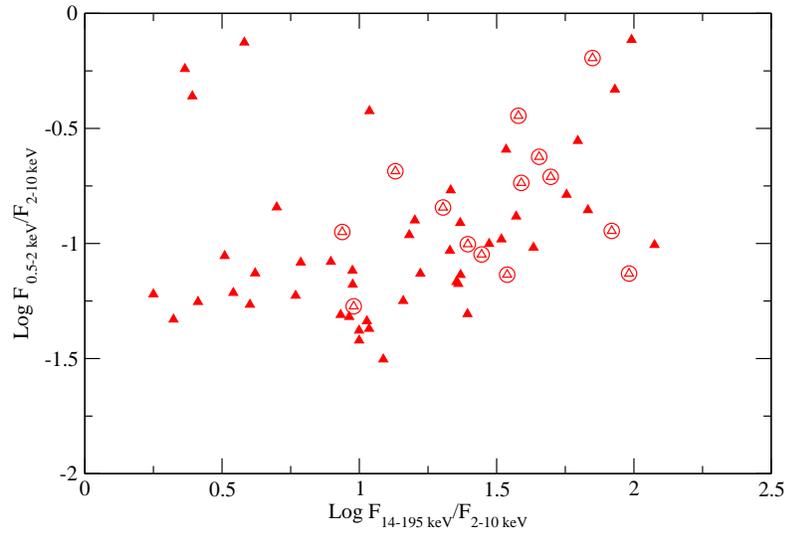}
\caption{Color-color diagram for the Seyfert 2 galaxies in our sample without published X-ray column density values (solid red triangles). Fluxes for the soft (0.5-2~keV) and medium (2-10~keV) X-ray   were taken from the  seven year {\it Swift}-XRT point source catalog \citep{2013A&A...551A.142D}. For comparison, we identified the Compton Thick sources (Table~\ref{CT}) as open red triangles-circles.     \label{HR}}
\end{figure}

As we mentioned before, the fraction of CT AGN in deep hard X-ray surveys  is   $17\% \pm 3\%$ \citep{2013arXiv1302.2451B}. This fraction of CT AGN is in agreement  with the value derived from the northern Galactic cap of the 58-month BAT catalog, where they found that up to 15\% of their sample could be CT \citep{2013ApJ...763..111V}. Since we identified only 18 CT candidates in our sample, this suggests  that there may be  some heavily obscured  Seyfert 2 galaxies that are unaccounted for.  In order to have an estimate of the X-ray column density for the  sources without published values (51 out of 157),    we used a simple  
color-color diagram  initially presented in \cite{2008ApJ...674..686W}, and later in \cite{2009ApJ...690.1322W}. Figure~\ref{HR} shows  the soft/medium (0.5-2~keV flux/2-10~keV flux) and  the hard/medium (14-195~keV flux/2-10~keV flux) ratios for the 51 Seyfert 2 galaxies without published X-ray column densities. Fluxes for the soft (0.5-2~keV) and medium (2-10~keV) X-ray   were taken from the  seven year {\it Swift}-XRT point source catalog \citep{2013A&A...551A.142D}. In the diagram we  also show the position for the CT sources in our sample (open red triangles-circles). It is clear from this comparison that our CT sources are located in  the same branch of heavily obscured sources ${\rm N_H > 10^{23}}$~${\rm cm^{-2}}$ which  extends towards the upper right side of the diagram \citep[see Figure~3 in ][]{2009ApJ...690.1322W}. We find that the vast majority of the CT sources have  hard/medium flux ratios, 14-195~keV flux/2-10~keV flux $>$ 10; thus this is a simple approximation for identifying   heavily obscured or CT candidates.  We apply this criteria to identify heavily obscured sources from our sample of Seyfert 2 galaxies, in particular,  those  sources without published column densities.  Overall, we find  26 sources with hard/medium ratios  greater than $\sim$ 10, indicative of  high column densities (see  Table~\ref{CC} for details). We performed a partial correlation analysis in our Seyfert 2 population after excluding the CT sources (Table~\ref{CT}) and our candidates of  heavily obscured Seyfert 2 galaxies without published X-ray column densities (Table~\ref{CC}). We find that  there is no statistically significant correlation between the 14-195 keV and FIR luminosities (70 and 160~$\micron$) in Seyfert 2 galaxies when the influence of  distance is excluded (see Table\ref{tablek}).  These results suggest that Compton scattering in the BAT band is not responsible for the lack of a correlation between the hard X-ray and FIR luminosities.


\begin{deluxetable}{lcc}
\tabletypesize{\scriptsize}
\tablewidth{0pt}
\tablecaption{Properties of Compton thick or  heavily obscured  Seyfert 2 galaxies, ${\rm N_H > 10^{24} cm^{-2}}$}
\tablehead{
\colhead{Name}           & \colhead{$N_H$}           & \colhead{Reference}\\   
                         & \colhead{${\rm 10^{24} cm^{-2}}$}           &  \\   
\tableline
\colhead{(1)} & \colhead{(2)} & \colhead{(3)} \\     
}
\startdata
CGCG 420-015              & 1.46 & (1) \\ 
ESO 005-G004              & 1.01 & (2)\\
ESO 137-G034              & $>$ 1.5& (3)\\
MCG-01-30-041             & 1.45 & (4) \\    
Mrk 3                     & 1.36 & (5) \\
Mrk 417                   & 1.20 & (4) \\  
NGC 424                   & 1.00 & (6) \\   
NGC 1365                  & 4.00 & (7) \\     
NGC 3079                  & 5.40 & (6) \\
NGC 3281                  & 2.00 & (8) \\     
NGC 3393                  & 4.50 & (6) \\         
NGC 4939                  & $>$10.00 & (9)  \\     
NGC 4941                  & 1.32 &  (4)\\    
NGC 5728                  & 1.39 & (10) \\     
NGC 6240                  & 1.83 & (6) \\      
NGC 6552                  & $>$1.00 & (11)\\     
NGC 7582                  & 1.10 & (6)\\        
UGC 05881                 & 2.45 &  (4)  \\

\enddata
\tablecomments{Column 1: galaxy name. Column 2: X-ray absorbing  column density. Column 3: references for the column density. (1) \cite{2011A&A...525A..38S}; (2) \cite{2007ApJ...664L..79U}; (3) \cite{2009MNRAS.394L.121M}; (4) \cite{2013ApJ...763..111V}; (5) \cite{2005MNRAS.360..380B};  (6) \citep{2011ApJ...728...58B}; (7) \cite{2009ApJ...705L...1R}; (8) \cite{2002A&A...381..834V}; (9) \cite{1998A&A...338..781M};  (10) \cite{2010ApJ...717..787C}; (11) \cite{1999ApJS..121..473B}} 
\label{CT}
\end{deluxetable}

On the other hand, if Seyfert 2 galaxies are intrinsically less luminous than Seyfert 1 galaxies, then the lack of a correlation  suggests that the FIR-X-ray connection  is luminosity dependent. In this scenario,  low luminosity AGN show little or  no relationship  between the AGN  and the  FIR luminosity, because  the latter is dominated by dust heated primarily by the stellar activity in the host galaxy \citep[e.g.,][]{2012A&A...545A..45R}. At higher luminosities (where the  AGN dominate), dust heated by the AGN becomes non-negligible, thereby, creating the  correlation between the intrinsic power of the AGN and the FIR \citep[e.g.,][]{2009MNRAS.399.1907N,2012A&A...545A..45R}.  

\begin{deluxetable}{lcc}
\tabletypesize{\scriptsize}
\tablewidth{0pt}
\tablecaption{Candidates for Compton thick or  heavily obscured  Seyfert 2 galaxies, ${\rm N_H > 10^{23} cm^{-2}}$}
\tablehead{
\colhead{Name}           & \colhead{${\rm F_{14-195 keV}/F_{2-10 keV}}$}   & \colhead{${\rm F_{0.5-2 keV}/F_{2-10 keV}}$}         \\   
\tableline
\colhead{(1)} & \colhead{(2)} & \colhead{(3)} \\

}
\startdata

2MASX J01064523+0638015        	&	10.66	&	0.05	\\
2MASX J01073963-1139117        	&	22.57	&	0.07	\\
2MASX J04234080+0408017        	&	32.92	&	0.10	\\
2MASX J06411806+3249313        	&	10.86	&	0.04    \\
2MASX J06561197-4919499        	&	56.87	&	0.16	\\
2MASX J07262635-3554214        	&	21.36	&	0.09	\\
2MASX J09360622-6548336        	&	15.92	&	0.13	\\
2MASX J20101740+4800214        	&	24.76	&	0.05	\\
2MFGC 02280                    	&	118.83	&	0.10	\\
CGCG 102-048                   	&	43.07	&	0.10	\\
CGCG 312-012                   	&	16.68	&	0.07	\\
ESO 244-IG030                  	&	21.50	&	0.17	\\
ESO 374-G044                   	&	14.45	&	0.06	\\
ESO 439-G009                   	&	34.23	&	0.26	\\
ESO 565-G019                   	&	23.37	&	0.07	\\
II Zw 083                       	&	15.20	&	0.11	\\
MCG-07-03-007                 	&	68.01	&	0.14	\\
MCG+02-21-013                 	&	29.70	&	0.10	\\
MCG+06-16-028                 	&	85.26	&	0.47	\\
MCG+06-49-019                 	&	10.88	&	0.38	\\
MCG+11-11-032                 	&	12.23	&	0.03	\\
NGC 1106                       	&	98.05	&	0.77	\\
NGC 1125                       	&	62.47	&	0.28	\\
SBS 0915+556                   	&	23.30	&	0.12	\\
UGC 01479                      	&	22.92	&	0.07	\\
VII Zw 073                      	&	37.26	&	0.13	\\

\enddata
\tablecomments{Column 1: galaxy name. Column 2: the soft/medium (0.5-2~keV flux/2-10~keV flux). Column 3:  the hard/medium (14-195~keV flux/2-10~keV flux). Fluxes for the soft (0.5-2~keV) and medium (2-10~keV) X-ray   were taken from the  seven year {\it Swift}-XRT point source catalog \citep{2013A&A...551A.142D}} 
\label{CC}
\end{deluxetable}

\section{The Far-Infrared Colors of the BAT sample }

\begin{figure}
\epsscale{.80}
\plotone{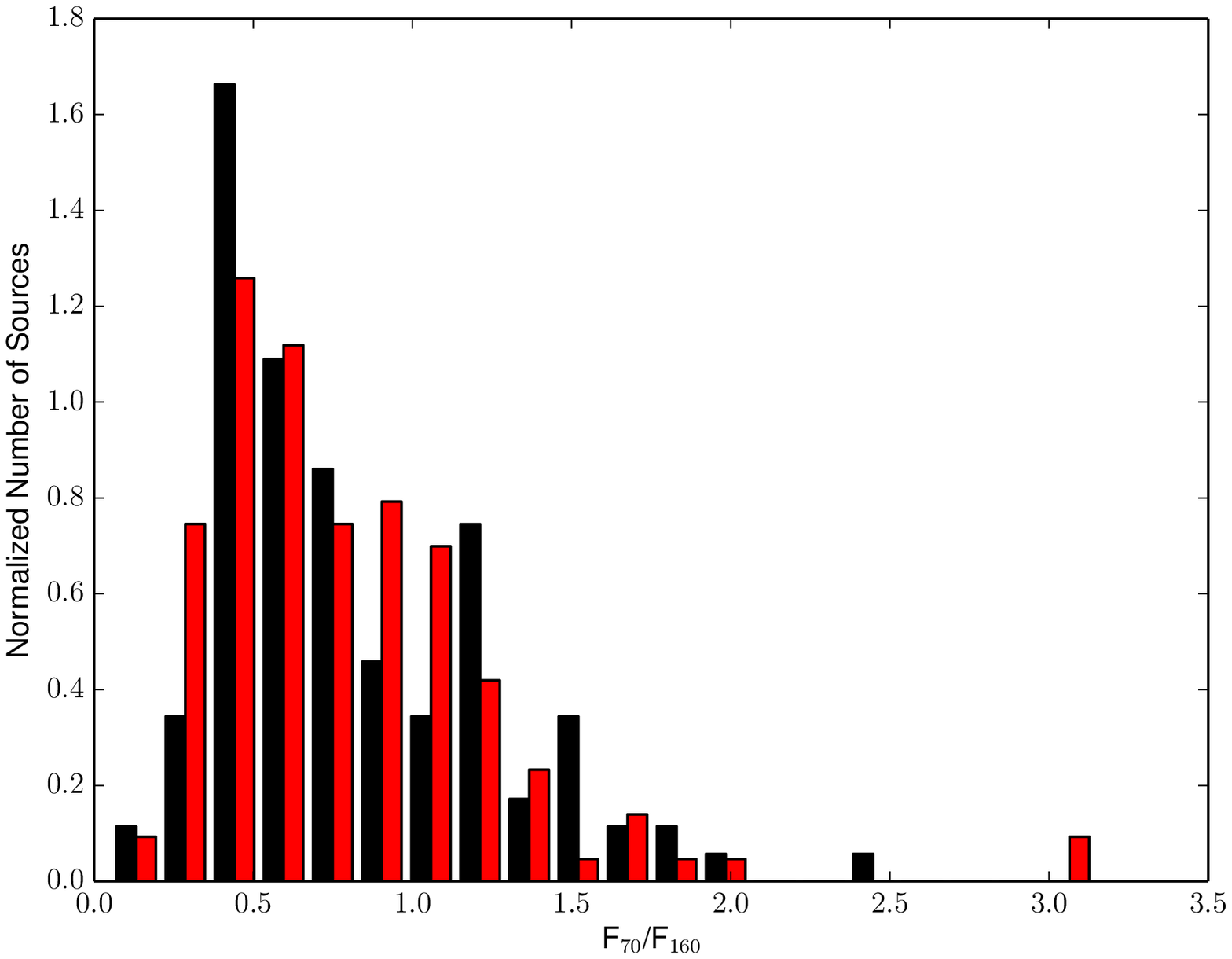}
\caption{The distribution of the F$_{70}$/F$_{160}$ ratio for Seyfert 1 and Seyfert 2 galaxies. The legend is the same as in  Figure~\ref{fig1}.  Upper limits are not included in the distribution.    \label{fig4}}
\end{figure}

\begin{figure}
\epsscale{.80}
\plotone{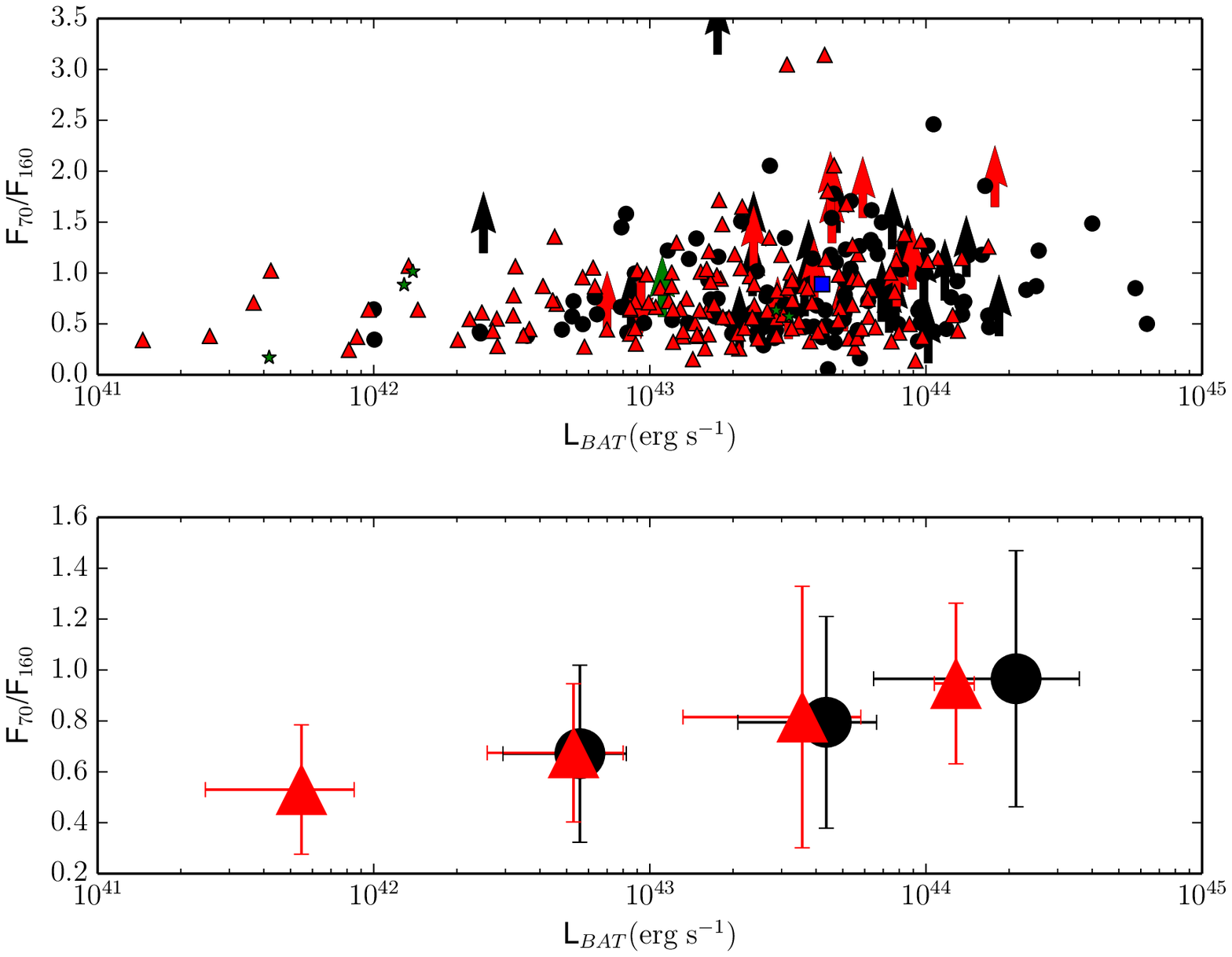}
\caption{Upper panel: correlation between the FIR color, F$_{70}$/F$_{160}$, and the 14-195 keV luminosity (L$_{BAT}$). Lower panel: binned correlation between the FIR color, F$_{70}$/F$_{160}$, and the 14-195 keV luminosity (L$_{BAT}$). We choose 4 equally spaced bins within the range of the 14-195 keV luminosity.  Symbols are: Seyfert 1 (black circles), Seyfert 2 (red triangles), LINERs (green stars) and AGN (blue square)\label{fig5}}
\end{figure}

Typically,  FIR emission from galaxies peakes  between $\sim$ 100 - 160~$\micron$. Thus, the  F$_{70}$/F$_{100}$ versus F$_{100}$/F$_{160}$  can be used to trace the peak of the galaxy SED \citep[e.g.,][]{2013A&A...557A..95R}.  However, in the absence of 100~$\micron$  observations the F$_{70}$/F$_{160}$ ratio may still provide   useful information regarding the peak of the FIR SED and the dust properties. Figure~\ref{fig4} shows the distribution of this ratios for  between Seyfert 1 and Seyfert 2 galaxies in the BAT sample. The  K-S test null probability is 53.5$\%$ which indicates that the two galaxy populations have similar FIR colors,  in agreement with the statistical similarities between the monochromatic luminosities at 70 and 160 $\micron$ (see Table~\ref{table2}).

Figure~\ref{fig5} shows  the  F$_{70}$/F$_{160}$ ratio versus the intrinsic X-ray luminosity of the AGN. We find  a trend of increasing F$_{70}$/F$_{160}$ with  BAT luminosity. Our analysis reveals a weak but statistically  significant correlation with a generalized Spearman rank-order correlation coefficient and probability (with censored data), $\rho=0.174$, $P_\rho=4.3\times 10^{-2}$ for Seyfert 1 galaxies  and a similar correlation for Seyfert 2 galaxies with $\rho=0.236$, $P_\rho=3.9\times 10^{-3}$.  This result implies  that there may be a dust grain distribution with  a warm (nuclear)  dust component, heated by the very energetic environment in the proximity of  the AGN (with a strong contribution to the 70~$\micron$ continuum), and a colder dust component  farther out in the outer regions of the AGN torus and/or  the host galaxy (the primary contributor to  the 160~$\micron$ continuum). In order to investigate  the AGN contribution to the FIR colors, we compared our observed colors with  predictions from three different  torus models:  the smooth, continuous  torus from \cite{2006MNRAS.366..767F}\footnote{http://users.ugent.be/~jfritz/jfhp/TORUS.html} and  the clumpy torus models from \cite{2010A&A...523A..27H}\footnote{http://www.sungrazer.org/CAT3D.html} and \cite{2012MNRAS.420.2756S} (``smooth", ``two-phase" and ``clumps-only")\footnote{https://sites.google.com/site/skirtorus/home}. We find that there is no region in parameter space (input parameters for the different models) that can reproduce  F$_{70}$/F$_{160}$ ratios less than unity.  In addition, the vast majority of the BAT AGN (74\% of all the sources with detections in both FIR bands) have F$_{70}$/F$_{160}$ ratios less than unity.

\begin{figure}
\epsscale{.80}
\plotone{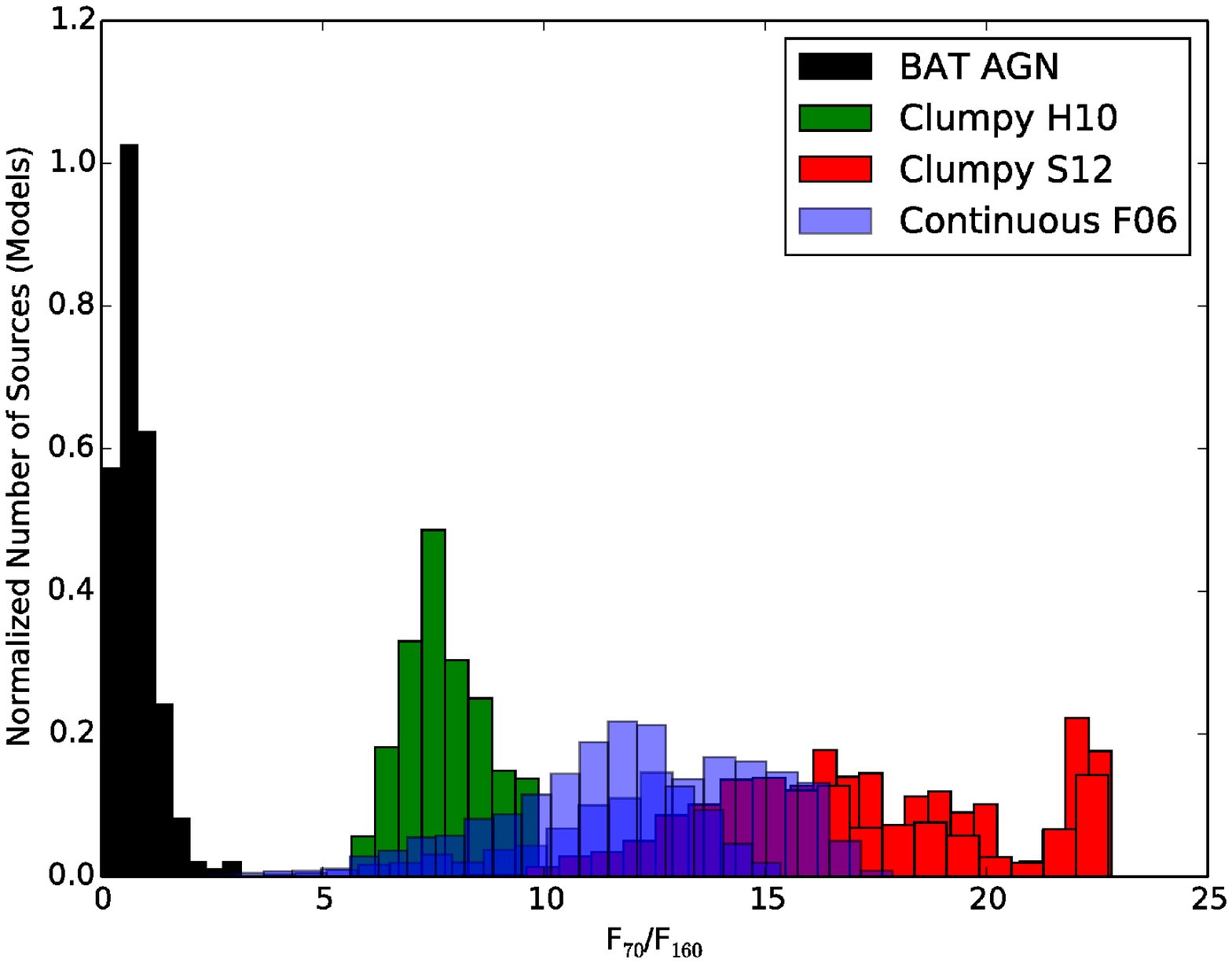}
\caption{A comparison between  the observed distribution of the FIR colors, F$_{70}$/F$_{160}$, for the BAT sample and predictions from different torus models. A smooth, continuous  torus model from \cite{2006MNRAS.366..767F} (F06) for two different torus opening angles,  $\Theta = 140\arcdeg$ and  $\Theta = 60\arcdeg$. Clumpy torus models from  \cite{2010A&A...523A..27H} (H10) and \cite{2012MNRAS.420.2756S} (S12) with a combination of  ``smooth", ``two-phase" and ``clumps-only" models. For the sake of simplicity, we combined, within each torus model,   all the predictions from  different input parameters.   The histogram is normalized by unit of area.  \label{comp_torus}}
\end{figure}

 All these models, on average, overpredict the  F$_{70}$/F$_{160}$ ratios  by at least a  factor of 2-3 \citep[][]{2006MNRAS.366..767F,2010A&A...523A..27H} and as much as a factor of $\sim$ 20 \citep[][]{2006MNRAS.366..767F,2012MNRAS.420.2756S} (see Figure~\ref{comp_torus}). Continuous models with a torus full-opening angle of $\Theta = 60\arcdeg$ \citep[covering factor, ][]{2006MNRAS.366..767F}  can reproduce  the most extreme sources in our sample  at the high end of the observed FIR color distribution with F$_{70}$/F$_{160}  > \sim 2.5$, namely Mrk~841, MCG-05-23-016  and ESO~103-035. Note that the observed BAT FIR SED is  the  combined effect from the AGN (non-thermal) and a stellar (thermal) radiation field. Therefore, it is not unexpected for the torus models to underpredict the observed FIR part of the SED without an extra stellar component (see discussion in Section~\ref{bat}). However,  the weak but probably  statistically significant correlation between the intrinsic power of the AGN and the FIR emission in Seyfert 1 galaxies suggests that there is a  subset of the  BAT AGN in which the FIR emission is dominated by the  non-stellar contribution (AGN). This population includes   AGN dominated sources with weak or no signatures of star formation at other wavelengths \citep[e.g., ][]{2014MNRAS.438..647H}. Therefore,  the previous results suggest that current  torus models may underpredict the intrinsic  AGN  emission at 160~$\micron$, relative to the 70~$\micron$, by having a steeper FIR SED.

\section{Comparison with Normal Galaxies}

\begin{figure}
\epsscale{.80}
\plotone{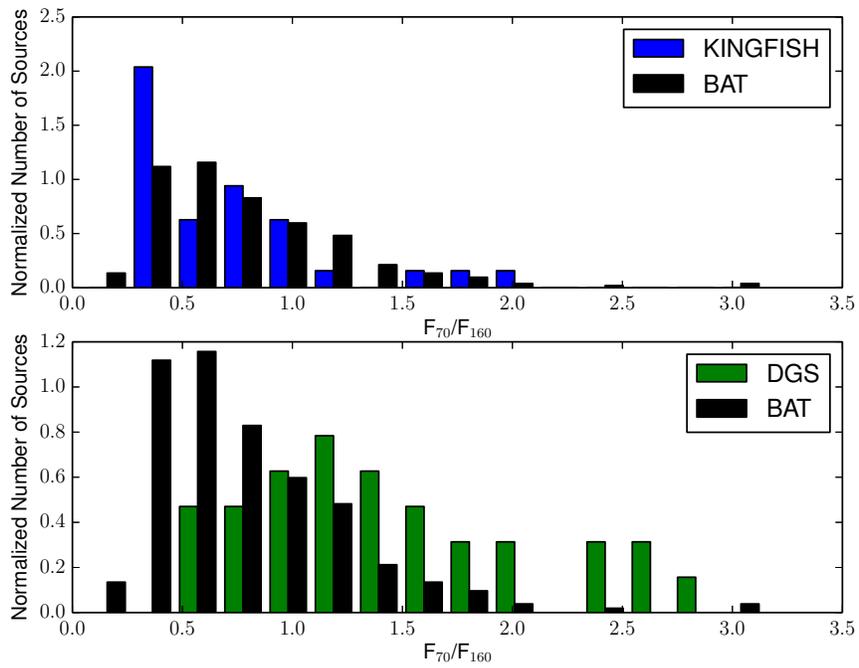}
\caption{A comparison between  the distributions of the F$_{70}$/F$_{160}$ ratios for the entire BAT sample with the   KINGFISH sample (upper panel) and the dwarf galaxy survey DGS (lower panel). The histograms are normalized by unit of area.   \label{fig10}}
\end{figure}

The  F$_{70}$/F$_{160}$ ratio  provides useful information when comparing with other samples of galaxies. Figure~\ref{fig10} shows the F$_{70}$/F$_{160}$ distribution  between the BAT AGN, the Key Insights on Nearby Galaxies: a Far-Infrared Survey with Herschel (KINGFISH) sample of nearby galaxies \citep{2012ApJ...745...95D} and the dwarf galaxy survey (DGS) \citep{2013A&A...557A..95R}. The KINGFISH sample  is an imaging and spectroscopic survey of 61 nearby (d $<$ 30 Mpc) galaxies draw mainly  from the Spitzer Infrared Nearby Galaxies Survey \citep[SINGS; see ][ for details]{2003PASP..115..928K}. The KINGFISH sample covers a wide range of galaxy properties and local interstellar medium  environments found in the nearby Universe. The {\it Herschel}  DGS is a photometric and spectroscopic sample of dwarf galaxies  in the local universe chosen to cover a wide range of physical conditions, including a wide range of metallicities \citep{2013PASP..125..600M}.

 For the KINGFISH sample, we selected only the ``normal" galaxies, e.g., the non-AGN galaxies as classified by their optical spectra \citep{2010ApJS..190..233M}.  Conversely, for the comparison  we included all BAT AGN. From Figure~\ref{fig10} it is clear that the BAT AGN and the KINGFISH normal galaxies span a similar range of values with a K-S null probability of 31$\%$, in other words, the BAT AGN FIR colors are statistically indistinguishibles from those in normal galaxies. On the other hand, the FIR colors for the BAT AGN are statistically different than those in the DGS sample with a K-S test null probability of 0.0 $\%$.  The majority of the BAT AGN and KINGFISH galaxies have F$_{70}$ $<$ F$_{160}$ corresponding to colder dust, whereas most of the dwarf galaxies peak at shorter wavelengths, F$_{70}$/F$_{160} > 1$, suggesting a warmer dust due to the harder radiation field illuminating the environment surrounding young stars. Note that  F$_{70}$/F$_{160}$ is also  sensitive to metallicity, in that the lower metallicity galaxies have  warmer dust than that found  in their metal-rich counterparts \citep{2013A&A...557A..95R}.

\section{Conclusions}
We present FIR  flux densities and maps for 313 hard X-ray selected AGN from the low-redshift ($z<0.05$) 58-month Swift-BAT survey in two PACS bands at 70 and 160~$\micron$. Of the 313 sources,   94\% and 83\% are detected in the FIR by  PACS at 70 and 160~$\micron$, respectively. From our analysis we find the following:

(1) Using the K-S test the FIR luminosity distributions at 70 and 160~$\micron$ for the Seyfert 1 and Seyfert 2 galaxies are indistinguishable from one another.  This result suggests that the FIR emission is isotropic, e.g.,  independent of orientation. We propose two different interpretations for the isotropic nature of the FIR emission: (i) star formation, which is isotropic, and/or (ii) some AGN torus models predict isotropy at FIR wavelengths. Regarding the former scenario,   if star formation dominates the FIR emission, then it  has to happen in a very compact, nuclear region, suggesting a connection between the AGN and the nuclear cold molecular gas \citep[see][]{2014ApJ...781L..34M}. In the latter scenario,  torus models predict a similar shape for the FIR part of the SED for both types of Seyferts \citep[e.g.,][]{2003PASP..115..928K,2012MNRAS.420.2756S}. However,  some torus  models predict an angle dependence on the FIR luminosity with  Seyfert 1 galaxies having higher luminosities than Seyfert 2 galaxies. For these models, star formation would also be needed \citep[e.g., ][]{2006MNRAS.366..767F}

Using the two-sample test for censored data, the 160~$\micron$  luminosity distributions between Seyfert galaxies may be distinguishable at a weak but possibly statistically significant level.  By including the non-detections, the Seyfert 2 galaxies are found to have a very slightly higher mean luminosity than the Seyfert 1 galaxies. If the FIR emission is a good probe of star formation, then this result suggests that Seyfert 2 galaxies have a slightly higher SFR than Seyfert 1 galaxies, which could imply  a time-dependent evolutionary  scenario, where the AGN are more obscured when the SFRs are at their higher values (see discussion in Section~\ref{corre}).

(2) A K-S test on the present data shows that the 14-195 keV luminosity distributions are statistically different between Seyfert 1 (unabsorbed) and Seyfert 2 (absorbed) galaxies. Our analysis suggests that the statistical differences between AGN types  are    fundamental,  with  absorbed AGN  intrinsically less luminous than  unabsorbed AGN.  This scenario  is supported by the difference in the luminosity function between  the two types of Seyferts for the BAT sample, with absorbed AGN having, on average,  lower luminosities than unabsorbed AGN \citep[e.g.,][]{2003ApJ...584L..57C,2003ApJ...596L..23S,2003ApJ...598..886U,2005AJ....129..578B,2005ApJ...635..864L,2005A&A...441..417H,2011ApJ...728...58B}.  Overall, this is a challenge  for the basic predictions from the unified model of AGN, where the intrinsic AGN luminosity should be independent of obscuration. Thus, one must  consider  a luminosity-dependent  unified model of AGN \citep[e.g.,][]{2005AJ....129..578B}.

(3) Using a non-parametric test for partial correlation with censored data, we find a statistically significant correlation between the AGN intrinsic power (BAT luminosity) and the FIR emission at 70 and 160 $\micron$ in Seyfert 1 galaxies. We find a better  correlation between the BAT and PACS luminosities at 70~$\micron$   than at 160~$\micron$. The correlation is also stronger for the most X-ray luminous objects.  The observed correlations  suggest two possible scenarios: (i) if we assume  that the FIR luminosity is a good tracer of star formation,  then there is a connection between star formation and the AGN at sub-kiloparsec scales,  and/or (ii) dust heated by the  AGN has a statistically significant  contribution to the FIR emission. Regarding the former scenario, and given the fact that the majority of the BAT sources have their FIR fluxes dominated by a point source located at the nucleus, one needs to consider that star formation  has to happen in a very compact, nuclear region \citep[][]{2014ApJ...781L..34M}. In the latter,   SFR indicators that rely on FIR emission, either through the individual infrared bands \citep[][]{2010ApJ...714.1256C} or the total FIR emission \citep[8--1000 $\micron$, ][]{1998ARA&A..36..189K}, need to consider the AGN contribution in their predictions.  However, all torus models studied  in the present work underpredict the amount of cold dust (from the outer regions of the torus) needed to match the observations; thus, a stellar component may still be required.

(4) A non-parametric test for partial correlation with censored data reveals that there is no statistically significant  correlation between the 14-195 keV and FIR luminosities in Seyfert 2 galaxies when the influence of the distance is excluded. The results presented in this paper  support a scenario in which   Seyfert 2 galaxies are intrinsically less luminous than Seyfert 1 galaxies suggesting that  the FIR-X-ray connection  is luminosity dependent. In other words, in   low luminosity AGN (Seyfert 2 galaxies), the FIR may be dominated by dust heated primarily by the stellar activity in the host galaxy \citep[e.g.,][]{2012A&A...545A..45R}, while at higher luminosities (Seyfert 1 galaxies)  the contribution from dust heated by the AGN  overwhelms the star formation contribution, thereby, creating the  correlation between the intrinsic power of the AGN and the FIR. 


(5) Using a K-S test, the distributions of the  F$_{70}$/F$_{160}$ ratios for the Seyfert 1 and 2 galaxies are indistinguishable from one another. They are also indistinguishable from those of normal star forming galaxies. In general, we find that the vast  majority of the BAT AGN (74\% of all  sources with a detection in both FIR bands) have F$_{70}$/F$_{160}$ ratios less than unity.  Assuming that the FIR is dominated by star formation,  these results are in agreement with the isotropic nature of the star formation emission.  However, one must consider that the weak but statistically significant correlation between the intrinsic power of the AGN and the FIR emission suggests that there must be an underlying population of BAT AGN in which the FIR emission is dominated by the  non-stellar contribution (the AGN torus). This population includes   AGN dominated sources with weak or no signatures of star formation at other wavelengths, e.g., Mrk 3 \citep[e.g.,][]{2014MNRAS.441..630S}. Nonetheless, when comparing with predictions from different torus models, we find that there is no region in parameter space (input parameters for the different models) that can reproduce  F$_{70}$/F$_{160}$ ratios less than unity. All these models, on average, overpredict the  F$_{70}$/F$_{160}$ ratios  by at least a  factor of 2-3 \citep[][]{2006MNRAS.366..767F,2010A&A...523A..27H} and by as much as a factor of $\sim$ 20 \citep[][]{2006MNRAS.366..767F,2012MNRAS.420.2756S}. Continuous models with a torus full-opening angle of $\Theta = 60\arcdeg$ \citep[covering factor, ][]{2006MNRAS.366..767F}  can only reproduce  the most extreme sources at the high end of the observed FIR colors distribution with F$_{70}$/F$_{160}  > \sim 2.5$, namely Mrk~841, MCG-05-23-016  and ESO~103-035. Thus, current  torus models may underpredict the AGN  emission at 160~$\micron$ (colder dust) relative to 70~$\micron$ with  a steeper FIR SED.

(6) Using a Spearman rank-order analysis, the 14-195 keV luminosities for the Seyfert 1 and 2 galaxies are weakly statistically correlated with the F$_{70}$/F$_{160}$  ratios.   This result suggests two possible interpretations: (i) a time-evolutionary sequence, as discussed in Section~\ref{corre}, where SFRs are higher in the more X-ray luminous sources, producing  a trend of increasing AGN luminosity and  FIR temperature, and/or (ii) the existence of a dust grain distribution with  a warm (nuclear)  dust component, heated by the very energetic environment in the proximity of  the AGN (with a strong contribution to the 70~$\micron$ continuum), and a colder dust component  farther out in the outer regions of the AGN torus and/or  the host galaxy (the primary contribution to  the 160~$\micron$ continuum).



\acknowledgments

We thank the referee for very useful comments that improved the paper. We would also like to acknowledge  Dr. Ranjan Vasudevan for valuable comments on the work. We gratefully acknowledge support from NASA through Herschel contracts 1436909 (R.F.M., M.M., T.T.S.), 1427288 (A.J.B.), and  1447980 (L.L.C.). PACS has been developed by a consortium of institutes led by MPE (Germany) and including UVIE (Austria); KU Leuven, CSL, IMEC (Belgium); CEA, LAM (France); MPIA (Germany); INAF-IFSI/OAA/OAP/OAT, LENS, SISSA (Italy); IAC (Spain). This development has been supported by the funding agencies BMVIT (Austria), ESA-PRODEX (Belgium), CEA/CNES (France), DLR (Germany), ASI/INAF (Italy), and CICYT/MCYT (Spain). The Herschel spacecraft was designed, built, tested, and launched under a contract to ESA managed by the Herschel/Planck Project team by an industrial consortium under the overall responsibility of the prime contractor Thales Alenia Space (Cannes), and including Astrium (Friedrichshafen) responsible for the payload module and for system testing at spacecraft level, Thales Alenia Space (Turin) responsible for the service module, and Astrium (Toulouse) responsible for the telescope, with in excess of a hundred subcontractors. This publication makes use of data products from the Wide-field Infrared Survey Explorer, which is a joint project of the University of California, Los Angeles, and the Jet Propulsion Laboratory/California Institute of Technology, funded by the National Aeronautics and Space Administration. The Cornell Atlas of Spitzer/IRS Sources (CASSIS) is a product of the Infrared Science Center at Cornell University, supported by NASA and JPL. This research has made use of NASA's Astrophysics Data System. This research has made use of the NASA/IPAC Extragalactic Database (NED) which is operated by the Jet Propulsion Laboratory, California Institute of Technology, under contract with the National Aeronautics and Space Administration.

\appendix

\section{{\it Herschel} PACS images}
\begin{figure}
\figurenum{10}
\plotone{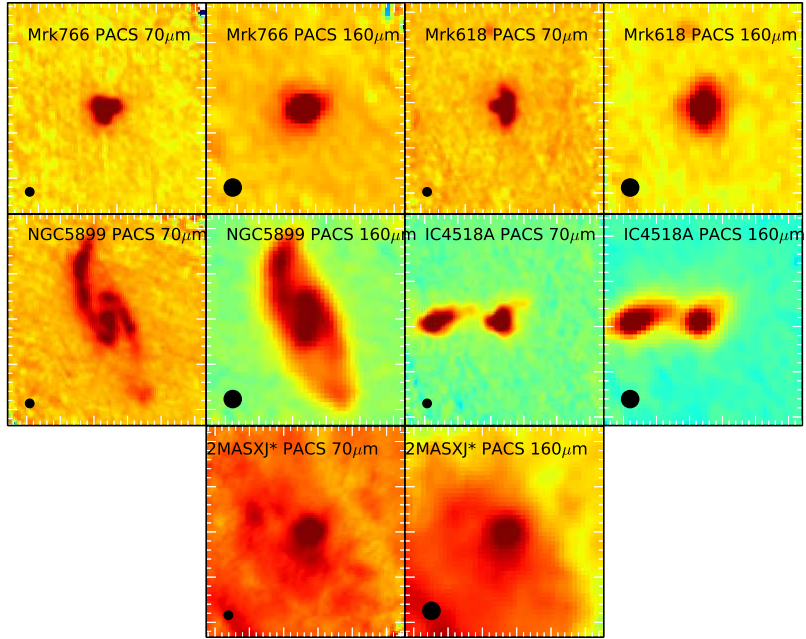}
\caption{PACS images for different types of morphologies for the BAT sample. The images were randomly selected to show an example of a  point-like source (Mrk 766), a slightly extended source (Mrk 618), a fully resolved source with complex structures (NGC 5899), a dual system (IC4518 pair) and a source contaminated  from Galactic cirrus (2MASXJ20183871+4041003).  The images are displayed with an inverse hyperbolic sine scaling \citep{1999AJ....118.1406L}. All the {\it Scanamorphos} images are presented in their native resolution and pixel size, 1.4 and 2.85\arcsec/pixel at 70 and 160\micron, respectively. The beam size at each wavelength is indicated by a black filled circle at the bottom left of each panel. The angular size of each panel is 2.2\arcmin$\times$2.2\arcmin.   North is up, and East is left for all the images. All the PACS images for the BAT sample are available in electronic form in the on-line version. \label{mosaic}}
\end{figure}

Figure~\ref{mosaic} shows the PACS {\it Scanamorphos}  images for the BAT sample. The images are displayed with an inverse hyperbolic sine scaling \citep{1999AJ....118.1406L}. All the images are presented in their native resolution and pixel size, 1.4 and 2.85\arcsec/pixel at 70 and 160\micron, respectively. The beam size at each wavelength is indicated by a black filled circle at the bottom left of each panel. The angular size of each panel is 2.2\arcmin$\times$2.2\arcmin.  North is up, and East is left for all the images. Figures 10.1–1.105 are available in the online version of the Journal.
 

\clearpage

\bibliographystyle{apj}

\bibliography{ms}
\clearpage



\end{document}